October 2021    **REVIEW OF THE CLASSIFICATION AND PROPERTIES**    Issue 1
              **OF 55 VARIABLE STARS IN SCUTUM**

CROZZA CLAUDIO[1], CURELAR SARA[1], DELL'AGLIO DEMETRIO[1],
LA SCALA FRANCESCA[1], MILLITARI ANGELO[1], MONTELLA ALESSANDRO[1], OROBELLO CHIARA[1]
BENNA CARLO[2], GARDIOL DANIELE[2] AND PETTITI GIUSEPPE[2]

1) IIS Curie Vittorini, Corso Allamano 130, 10095, Grugliasco (TO), Italy, TOIS03400P@istruzione.it

2) INAF-Osservatorio Astrofisico di Torino, via Osservatorio 20, I-10025 Pino Torinese (TO), Italy, giuseppe.pettiti@inaf.it

**Abstract**. This study aims to assess the properties and classification of 55 variable stars in Scutum, little studied since their discovery and reported in the Information Bulletin on Variable Stars (IBVS) 985 and update.
Using data from previous studies and several astronomical databases, we performed our analysis mainly utilizing a period analysis software and comparing the photometric characteristics of the variables in a Colour-Absolute Magnitude Diagram.
For all stars, the variability is confirmed. We discovered new significant results for the period and/or type of 17 variables and highlighted incorrect cross-reference names on astronomical databases for 3 stars. This assessment also identifies 12 cases for which the results from the ASAS-SN Catalog of Variable Stars are systematically not consistent with the original light curves.

1. **Introduction**

This article is a continuation of a previous work (Aglì et al. 2021) conducted at the Astrophysical Observatory of Turin on a group of variable stars originally characterized by IBVS 985 (Maffei, 1975).
The IBVS 985 provides the results of a systematic photographic search of variable stars in a field centered at R.A. = $18^h\ 14^m$, Dec. = $-14°\ 50'$ (1900.0), containing the objects M16 and M17, performed in blue and infrared band at the Astrophysical Observatory of Asiago beginning from the summer of 1967. This photographic survey led to the discovery of 197 new variables, mostly late spectral type stars with low effective temperature and maximum luminosity at infrared wavelengths, confirming the strong increase of the number of Mira type stars when the observations are made in the infrared band.
An update of the characteristics of 176 Mira and Semiregular variable stars originally observed and in particular their revised coordinates and finding charts were provided with a later paper (Maffei and Tosti, 2013). The update made use of discontinuous observations performed with the two Schmidt telescopes of the Astrophysical Observatory of Asiago and with the Schmidt telescope of the Catania Astrophysical Observatory from 1961 to 1991, with photographic emulsion 103a-O without or with filter GG13 and hypersensitized with $NH_3$ or pre-flashed I-N+RG5. A revision of variables identification which provides 2MASS counterparts for 19 stars was issued in 2018 (Nesci, 2018).

Initial classification and properties of variable stars discovered through surveys are sometimes affected by mistakes or uncertainties that remain undiscovered or unresolved due to the lack of new observations or further analysis. The variable stars of IBVS 985 are distributed in the constellations of Sagittarius, Serpens and Scutum and are generally faint and little studied since their discovery.

The previous paper (Aglì et al. 2021) was focused on variable stars located in the Sagittarius constellation. In this study we assess the original physical and photometric characteristics and the classification of the 55 variable stars of IBVS 985 belonging to the constellation of Scutum, based on data mined by more recent and public astronomical articles and databases. The provisional name, original coordinates, final designation, and main characteristics of the stars covered by this study are shown in Table 1.

In this paper, when the term 'original' is used for the characteristics of the variables, we refer to the updated data defined in the Atlas of 176 Mira/SR Light Curves (Maffei and Tosti, 2013).

Section 2 describes the method of our assessment. In section 2.1, we provide a general assessment of the classification type of the variables based on their Gaia photometric characteristics; a more detailed analysis for each variable, based on the results of previous studies and of our photometric analysis, is provided in section 2.2. Our conclusions are summarized in section 3.

Table 1    Original main characteristics of the variable stars in Scutum

| Provisional name | Original coordinates R.A. - Decl. (1950.0) | Variable Designation | Type | Mag. Range (I-N + RG5) | Period (d) |
|---|---|---|---|---|---|
| M010 | 18 22 30  -15 33.7 | V379 Sct | M | 13.0 ÷ 16.6 | 413.0 |
| M015 | 18 26 31  -15 10.0 | V405 Sct | M | 14.3 ÷ 16.2 | 307.0 |
| M024 | 18 27 06  -15 20.7 | NSV 10899 | I | 14.0 ÷ 15.8 | --- |
| M026 | 18 24 23  -15 51.5 | V386 Sct | M | 13.1 ÷ 17.1 | 426.0 |
| M061 | 18 23 44  -15 52.5 | V383 Sct | M | 14.2 ÷ 17.3 | 289.0 |
| M063 | 18 25 23  -15 34.3 | V387 Sct | M | 13.8 ÷ 16.4 | 379.0 |
| M064 | 18 25 14  -15 56.7 | V391 Sct | IS: | 13.4 ÷ 17.0 | --- |
| M065 | 18 26 20  -15 39.7 | V404 Sct | M | 15.5 ÷ 17.2 | 288.0 |
| M066 | 18 25 27  -15 27.8 | V394 Sct | M | 12.9 ÷ 15.9 | 314.0 |
| M067 | 18 25 50  -15 24.9 | V400 Sct | M | 12.3 ÷ 15.3 | 295.0 |
| M068 | 18 25 39  -15 21.4 | V397 Sct | M | 13.7 ÷ 16.4 | 312.0 |
| M069 | 18 23 17  -14 49.1 | V382 Sct | E: | 11.9 ÷ 14.7 | 592: |
| M071 | 18 22 09  -14 19.9 | V377 Sct | M | 12.2 ÷ 16.2 | 522.0 |
| M072 | 18 24 08  -14 43.7 | V385 Sct | M | 7.6 ÷ 11.6 | 397.0 |
| M073 | 18 25 05  -14 50.6 | V390 Sct | M | 12.5 ÷ 17.0 | 319.0 |
| M074 | 18 25 29  -15 08.0 | V395 Sct | M | 12.8 ÷ 16.7 | 310.0 |
| M075 | 18 25 44  -14 58.3 | V399 Sct | M | 13.2 ÷ 16.0 | 317.0 |
| M076 | 18 27 59  -15 39.6 | V421 Sct | M: | 12.8 ÷ 16.2 | 219.9 |
| M077 | 18 28 44  -15 29.1 | V424 Sct | M | 10.0 ÷ 15.5 | 474.0 |
| M078 | 18 28 33  -15 20.5 | V422 Sct | M | 12.1 ÷ 15.8 | 423.0 |
| M079 | 18 26 10  -14 45.9 | V402 Sct | M | 11.9 ÷ 15.4 | 312.0 |
| M080 | 18 26 03  -14 31.3 | V401 Sct | M | 11.8 ÷ 15.4 | 474.0 |

| Provisional name | Original coordinates R.A. - Decl. (1950.0) | Variable Designation | Type | Mag. Range (I-N + RG5) | Period (d) |
|---|---|---|---|---|---|
| M083 | 18 21 23  -13 17.6 | NSV 10736 | M | 14.0 ÷ 16.1 | 507.0 |
| M084 | 18 27 23  -14 27.4 | V415 Sct | M | 11.4 ÷ 15.2 | 303.0 |
| M085 | 18 27 53  -14 54.9 | V420 Sct | M | 11.3 ÷ 14.1 | 261.0 |
| M087 | 18 25 00  -13 45.0 | NSV 10832 | E | 12.8 ÷ 13.9 | --- |
| M128 | 18 24 02  -15 42.3 | V384 Sct | M: | 14.1 ÷ 17.4 | 454.0 |
| M134 | 18 26 31  -15 49.5 | V406 Sct | M: | 11.5 ÷ 15.4 | 447.0 |
| M135 | 18 26 40  -15 50.7 | V407 Sct | M | 11.6 ÷ 15.4 | 238.5 |
| M136 | 18 27 10  -15 30.6 | V413 Sct | M | 15.0 ÷ 17.0 | 333.0 |
| M137 | 18 27 22  -15 29.5 | V414 Sct | M | 14.4 ÷ 17.2 | 529.0 |
| M138 | 18 22 49  -14 44.8 | V380 Sct | M | 12.7 ÷ 15.5 | 447.0 |
| M145 | 18 25 42  -14 58.5 | V398 Sct | M | 13.4 ÷ 16.0 | 368.0 |
| M146 | 18 25 26  -14 14.3 | V393 Sct | M | 10.5 ÷ 15.2 | 381.5 |
| M147 | 18 20 27  -13 03.2 | V375 Sct | M | 14.4 ÷ 16.4 | 495.0 |
| M149 | 18 21 34  -12 53.8 | V376 Sct | M | 10.4 ÷ <16.0 | 590.0 |
| M150 | 18 25 21  -13 46.5 | NSV 10848 | Nova? | 14.4 ÷ <18.0 | --- |
| M151 | 18 26 50  -14 02.3 | V409 Sct | M | 12.2 ÷ 16.0 | 469.0 |
| M152 | 18 27 08  -14 12.3 | V412 Sct | M | 12.5 ÷ 16.0 | 408.0 |
| M153 | 18 27 46  -14 18.5 | V419 Sct | SRa | 14.0 ÷ 15.6 | 359.0 |
| M154 | 18 27 38  -14 23.7 | V418 Sct | M | 12.9 ÷ 15.4 | 411.0 |
| M187 | 18 22 10  -15 42.8 | V378 Sct | M | 13.0 ÷ 16.5 | 435.0 |
| M188 | 18 23 00  -15 33.9 | V381 Sct | SRa | 14.3 ÷ 16.7 | 417.0 |
| M192 | 18 25 37  -15 08.4 | V396 Sct | M | 13.8 ÷ <16.8 | 398.0 |
| M193 | 18 25 21  -15 16.6 | V392 Sct | M | 13.7 ÷ 16.5 | 480.0 |
| M198 | 18 26 11  -14 48.9 | V403 Sct | M | 12.5 ÷ 16.1 | 337.0 |
| M199 | 18 27 02  -14 59.9 | V410 Sct | M | 14.2 ÷ 15.8 | 327.0 |
| M200 | 18 28 50  -15 14.4 | V425 Sct | M | 12.0 ÷ 16.0 | 445.0 |
| M201 | 18 26 48  -14 48.2 | V408 Sct | M | 14.4 ÷ 16.4 | 321.0 |
| M202 | 18 27 25  -14 33.5 | V417 Sct | M | 11.0 ÷ 14.9 | 407.0 |
| M203 | 18 20 21  -12 39.5 | V374 Sct | E | 14.5 ÷ 16.1 | 408.0 |
| M204 | 18 24 45  -13 40.7 | V388 Sct | M | 12.3 ÷ 15.6 | 379.0 |
| M205 | 18 27 24  -14 23.6 | V416 Sct | M | 12.9 ÷ 15.5 | 300.0 |
| M206 | 18 28 34  -14 46.0 | V423 Sct | M | 12.7 ÷ 16.0 | 449.0 |
| M208 | 18 27 06  -14 44.9 | V411 Sct | SR | 12.9 ÷ 15.0 | 457.0 |

2. Data analysis

The data analysis was performed with the same approach adopted in the previous study dealing with variable stars belonging to Sagittarius constellation (Aglì et al. 2021).
For each variable star listed in Table 1, we first checked the Gaia source identifier reported by the SIMBAD database or, when this information was missing, we identified a Gaia counterpart using the variable coordinates derived from the original finding charts. If more than one Gaia identifier was possible, due to the uncertainty on the variable known position, we selected the Gaia source based on its expected photometric characteristics. A rationale of the selection criterion for the variable stars for which such approach was necessary is provided in the detailed data analysis of section 2.2.

Once the identity of all variable stars has been confirmed, in addition to the information available on SIMBAD, we looked for physical, spectroscopic, and photometric data available in other astronomical databases and catalogues, specifically the ones from the American Association of Variable Stars Observers (AAVSO, Kafka 2020), the All-Sky Automated Survey for Supernovae (ASAS-SN, Kochanek et al. 2017; Jayasinghe et al. 2019), the 2MASS All Sky Catalogue (Cutri et al. 2003), the Gaia Data Release 2 (Gaia Collaboration et al. 2016; Gaia Collaboration et al. 2018b), and the Gaia Data Early Release 3 (Gaia Collaboration et al. 2020a).

In Table 2 we list, for each variable, the Gaia DR2/EDR3 source ID and the apparent median Gaia G magnitude in DR2 and EDR3 with associated errors, because photometric system for G, Bp and Rp magnitudes in Gaia EDR3 is different from the one as used in Gaia DR2.

Table 3 shows the absolute magnitudes calculated using the distances available from the Gaia Distances to 1.33 billion stars in Gaia DR2 catalogue (Bailer-Jones et al. 2018) and also summarizes the effective temperature $T_{eff}$ and the Gaia colour index Bp-Rp. The interstellar extinction or circumstellar extinction was not considered in the absolute magnitude calculation because the extinction value estimation in the filter G is available only for 16 of our sample of 55 variable stars.

Table 2     Gaia DR2/EDR3 source ID and apparent median G magnitudes

| Name | Gaia DR2/EDR3 Source ID | Gaia DR2 Median G (mag.) | ΔG (±) | Gaia EDR3 Median G (mag.) | ΔG (±) |
|---|---|---|---|---|---|
| NSV 10736 | 4152543642155820032 | 16.151 | 0.087 | 16.215 | 0.092 |
| NSV 10832 | 4104438148705312640 | 13.056 | 0.019 | 13.122 | 0.015 |
| NSV 10848 | --- | --- | --- | --- | --- |
| NSV 10899 | 4103942582578175744 | 14.603 | 0.027 | 14.588 | 0.022 |
| V374 Sct | 4152798969359304320 | 14.767 | 0.040 | 14.984 | 0.035 |
| V375 Sct | 4152750693951525376 | 15.721 | 0.111 | 15.556 | 0.065 |
| V376 Sct | 4152590749356596864 | 14.055 | 0.074 | 14.430 | 0.066 |
| V377 Sct | 4104298274529689344 | 14.448 | 0.066 | 14.568 | 0.054 |
| V378 Sct | 4097936603350502912 | 15.119 | 0.058 | 15.025 | 0.027 |
| V379 Sct | 4097965740393789568 | 14.210 | 0.045 | 14.325 | 0.048 |
| V380 Sct | 4104201998517789312 | 15.137 | 0.054 | 15.041 | 0.035 |
| V381 Sct | 4097962201356698240 | 14.390 | 0.034 | 14.488 | 0.033 |
| V382 Sct | 4104199043611512192 | 13.361 | 0.043 | 13.311 | 0.024 |
| V383 Sct | 4097895509099999360 | 15.014 | 0.061 | 14.953 | 0.045 |
| V384 Sct | 4097903824157602560 | 15.518 | 0.100 | 15.644 | 0.079 |
| V385 Sct | 4104222614393218816 | 11.048 | 0.069 | 11.005 | 0.041 |
| V386 Sct | 4097898532723434368 | 14.916 | 0.075 | 14.700 | 0.065 |
| V387 Sct | 4097918220888613760 | 15.257 | 0.031 | 15.280 | 0.024 |
| V388 Sct | 4104451175395862784 | 14.690 | 0.108 | 14.688 | 0.076 |

| Name | Gaia DR2/EDR3 Source ID | Gaia DR2 Median G (mag.) | ΔG (±) | Gaia EDR3 Median G (mag.) | ΔG (±) |
|---|---|---|---|---|---|
| V390 Sct | 4104031058914107392 | 15.064 | 0.058 | 15.075 | 0.059 |
| V391 Sct | 4097142721628103424 | 12.952 | 0.005 | 12.947 | 0.006 |
| V392 Sct | 4103935229609066112 | 15.444 | 0.062 | 15.140 | 0.054 |
| V393 Sct | 4104335795346607872 | 13.760 | 0.079 | 13.849 | 0.039 |
| V394 Sct | 4103919454096002176 | 14.508 | 0.069 | 14.552 | 0.059 |
| V395 Sct | 4103963645044137088 | 14.397 | 0.071 | 14.414 | 0.059 |
| V396 Sct | 4103962854825729408 | 16.286 | 0.035 | --- | --- |
| V397 Sct | 4103921996717411712 | 14.998 | 0.054 | 15.084 | 0.053 |
| V398 Sct | 4104013741603507840 | 14.656 | 0.044 | 14.506 | 0.033 |
| V399 Sct | 4104013771581820416 | 14.778 | 0.057 | 14.981 | 0.070 |
| V400 Sct | 4103919351017148160 | 13.292 | 0.023 | 13.328 | 0.021 |
| V401 Sct | 4104137432283390464 | 13.598 | 0.037 | 13.696 | 0.032 |
| V402 Sct | 4104026660866978560 | 13.727 | 0.103 | 14.143 | 0.076 |
| V403 Sct | 4104023018734396800 | 14.794 | 0.063 | 14.933 | 0.042 |
| V404 Sct | 4103161894951509504 | 16.033 | 0.061 | 15.566 | 0.051 |
| V405 Sct | 4103948939130077056 | 15.709 | 0.069 | 15.511 | 0.048 |
| V406 Sct | 4103139011331374720 | 13.359 | 0.050 | 13.354 | 0.030 |
| V407 Sct | 4103138148004853248 | 13.150 | 0.032 | 12.968 | 0.035 |
| V408 Sct | 4104069468721467136 | 15.881 | 0.056 | 15.833 | 0.085 |
| V409 Sct | 4104366341161294976 | 15.028 | 0.118 | 15.267 | 0.050 |
| V410 Sct | 4103969348817568896 | 14.958 | 0.048 | 14.902 | 0.034 |
| V411 Sct | 4104437912536530432 | 13.784 | 0.052 | 13.605 | 0.039 |
| V412 Sct | 4104174304682462976 | 15.039 | 0.053 | 15.140 | 0.055 |
| V413 Sct | 4103185397013474432 | 15.501 | 0.048 | 15.526 | 0.035 |
| V414 Sct | 4103185637531623552 | 16.604 | 0.085 | 16.117 | 0.074 |
| V415 Sct | 4104132140880936576 | 13.859 | 0.039 | 13.853 | 0.047 |
| V416 Sct | 4104133519686449920 | 14.130 | 0.061 | 13.882 | 0.042 |
| V417 Sct | 4104126815120395776 | 13.223 | 0.029 | 13.339 | 0.036 |
| V418 Sct | 4104156197098012288 | 14.039 | 0.058 | 13.929 | 0.043 |
| V419 Sct | 4104157915085065088 | 15.267 | 0.039 | 15.274 | 0.038 |
| V420 Sct | 4104053633283796736 | 13.237 | 0.049 | 13.116 | 0.046 |
| V421 Sct | 4103167633028466816 | 14.142 | 0.052 | 14.046 | 0.045 |
| V422 Sct | 4103183266710080896 | 14.556 | 0.081 | 14.792 | 0.060 |
| V423 Sct | 4104062463737773568 | 14.743 | 0.055 | 14.623 | 0.051 |
| V424 Sct | 4103177047597293696 | 14.631 | 0.075 | 14.400 | 0.065 |
| V425 Sct | 4103289674449169408 | 14.356 | 0.048 | 14.363 | 0.049 |

Table 3    Absolute Magnitude, Effective Temperature and color index

| Name | DR2 Absolute Magnitude ($M_G$) | $\Delta M_G$ (±) | Teff (K) | $\Delta$ Teff (± K) | DR2 Bp-Rp |
|---|---|---|---|---|---|
| NSV 10736 | 5.65 | -1.69 ÷ 1.00 | 3281 | -6 ÷ 12 | 6.148 |
| NSV 10832 | -0.17 | -1.08 ÷ 1.11 | 3284 | -8 ÷ 59 | 6.515 |
| NSV 10848 | --- | --- | --- | --- | --- |
| NSV 10899 | 1.76 | -1.10 ÷ 0.91 | 3296 | -13 ÷ 97 | 4.771 |
| V374 Sct | 0.84 | -0.88 ÷ 0.91 | 3284 | -9 ÷ 60 | 6.919 |
| V375 Sct | 3.68 | -1.43 ÷ 1.17 | --- | --- | --- |
| V376 Sct | 1.57 | -1.31 ÷ 1.26 | 3281 | -5 ÷ 12 | 6.797 |
| V377 Sct | 1.88 | -1.31 ÷ 1.27 | 3280 | -5 ÷ 12 | 7.749 |
| V378 Sct | 3.28 | -1.58 ÷ 1.23 | 3281 | -5 ÷ 13 | 6.161 |
| V379 Sct | 0.22 | -0.87 ÷ 0.88 | 3284 | -8 ÷ 59 | 6.973 |
| V380 Sct | 4.29 | -1.28 ÷ 0.84 | 3279 | -11 ÷ 48 | 5.832 |
| V381 Sct | 1.12 | -1.08 ÷ 1.13 | 3284 | -9 ÷ 57 | 6.681 |
| V382 Sct | -0.76 | -0.85 ÷ 0.88 | 3281 | -5 ÷ 12 | 6.799 |
| V383 Sct | 2.62 | -1.30 ÷ 1.07 | 3284 | -9 ÷ 57 | 6.860 |
| V384 Sct | 2.09 | -1.03 ÷ 1.01 | 3283 | -8 ÷ 57 | 6.243 |
| V385 Sct | 0.17 | -1.13 ÷ 0.77 | 3294 | -15 ÷ 32 | 5.445 |
| V386 Sct | 1.58 | -1.07 ÷ 1.07 | 3284 | -8 ÷ 66 | 6.690 |
| V387 Sct | 3.50 | -1.39 ÷ 0.98 | 3282 | -11 ÷ 61 | 5.783 |
| V388 Sct | 0.61 | -0.86 ÷ 0.88 | 3285 | -7 ÷ 50 | 5.624 |
| V390 Sct | 2.15 | -1.20 ÷ 1.18 | 3284 | -8 ÷ 59 | 6.309 |
| V391 Sct | -1.10 | -0.72 ÷ 0.63 | 4078 | -185 ÷ 393 | 1.987 |
| V392 Sct | 5.00 | -0.84 ÷ 0.61 | 3284 | -9 ÷ 48 | 5.702 |
| V393 Sct | 1.75 | -1.53 ÷ 1.25 | 3281 | -6 ÷ 12 | 6.499 |
| V394 Sct | 2.36 | -1.16 ÷ 0.86 | 3284 | -16 ÷ 56 | 5.499 |
| V395 Sct | 1.54 | -1.19 ÷ 1.10 | 3283 | -10 ÷ 62 | 5.799 |
| V396 Sct | 2.96 | -1.08 ÷ 1.08 | 3287 | -15 ÷ 77 | 5.568 |
| V397 Sct | 3.38 | -1.25 ÷ 0.87 | 3284 | -16 ÷ 55 | 5.531 |
| V398 Sct | 1.05 | -0.97 ÷ 0.95 | 3296 | -8 ÷ 14 | 4.573 |
| V399 Sct | 2.02 | -1.27 ÷ 1.23 | 3287 | -12 ÷ 77 | 5.593 |
| V400 Sct | 1.04 | -1.23 ÷ 0.94 | 3281 | -11 ÷ 62 | 5.775 |
| V401 Sct | 1.04 | -1.24 ÷ 1.05 | 3284 | -8 ÷ 59 | 6.648 |
| V402 Sct | 0.37 | -1.06 ÷ 1.08 | 3284 | -9 ÷ 124 | 5.246 |
| V403 Sct | 1.24 | -1.02 ÷ 1.06 | 3284 | -9 ÷ 60 | 6.061 |
| V404 Sct | 2.20 | -0.91 ÷ 0.90 | 3380 | -101 ÷ 272 | 4.607 |
| V405 Sct | 2.48 | -1.11 ÷ 1.10 | 3304 | -24 ÷ 102 | 5.337 |
| V406 Sct | 3.24 | -0.67 ÷ 0.52 | 3284 | -8 ÷ 17 | 6.055 |
| V407 Sct | -0.47 | -0.94 ÷ 0.88 | 3281 | -6 ÷ 13 | 5.996 |

| Name | DR2 Absolute Magnitude ($M_G$) | $\Delta M_G$ ($\pm$) | Teff (K) | $\Delta$ Teff ($\pm$ K) | DR2 Bp-Rp |
|---|---|---|---|---|---|
| V408 Sct | 4.41 | -1.75 ÷ 1.21 | 3289 | -14 ÷ 124 | 4.832 |
| V409 Sct | 3.35 | -1.66 ÷ 1.22 | 3284 | -9 ÷ 43 | 5.669 |
| V410 Sct | 1.06 | -0.93 ÷ 0.96 | 3287 | -12 ÷ 77 | 5.607 |
| V411 Sct | -0.20 | -0.89 ÷ 0.93 | 3284 | -8 ÷ 60 | 6.569 |
| V412 Sct | 1.99 | -1.16 ÷ 1.17 | 3284 | -8 ÷ 47 | 6.030 |
| V413 Sct | 2.36 | -1.16 ÷ 1.18 | 3283 | -13 ÷ 53 | 6.029 |
| V414 Sct | --- | --- | --- | --- | --- |
| V415 Sct | 3.45 | -0.71 ÷ 0.54 | 3284 | -9 ÷ 60 | 5.895 |
| V416 Sct | 0.43 | -0.98 ÷ 1.01 | 3283 | -13 ÷ 44 | 5.925 |
| V417 Sct | 0.56 | -1.30 ÷ 1.24 | 3284 | -9 ÷ 60 | 6.229 |
| V418 Sct | -0.38 | -0.79 ÷ 0.81 | 3282 | -6 ÷ 17 | 6.159 |
| V419 Sct | 2.35 | -1.20 ÷ 1.16 | 3283 | -11 ÷ 63 | 5.811 |
| V420 Sct | 0.33 | -1.15 ÷ 0.99 | 3289 | -8 ÷ 87 | 5.515 |
| V421 Sct | 0.37 | -0.95 ÷ 0.95 | 3280 | -12 ÷ 51 | 5.763 |
| V422 Sct | 0.56 | -0.92 ÷ 0.95 | 3285 | -14 ÷ 73 | 5.569 |
| V423 Sct | 2.07 | -1.30 ÷ 1.18 | 3304 | -25 ÷ 103 | 5.321 |
| V424 Sct | 1.30 | -1.10 ÷ 1.13 | 3304 | -26 ÷ 119 | 5.271 |
| V425 Sct | 1.47 | -1.26 ÷ 1.28 | 3282 | -9 ÷ 62 | 7.014 |

Based on their Gaia EDR3 equatorial coordinates (transformed at J2000.0), we searched for each variable the photometric measurements and light curves analysis available from ASAS-SN database. If no variable is reported by ASAS-SN database we performed a period analysis of the photometric data of the source that matches the Gaia EDR3 equatorial coordinates. We analyzed the valid photometric data using version 2.51 and 2.60 of the light curve and period analysis software PERANSO (Paunzen and Vanmunster, 2016). The analysis of the period was performed using two methods: Lomb-Scargle (Lomb 1976, Scargle 1982) and ANOVA (Schwarzenberg-Czerny A., 1996), that are deemed powerful in detecting weak periodic signals, improving sensitivity of peak detection and damping alias periods. For each value of the period, reported in this paper, a Fisher Randomization Test, with 200 permutations, was run with PERANSO software, in order to confirm the significance of the period we found. All results reported in this paper have a False Alarm Probabilities (FAP) less than 1%, indicating a high significance of the result. Any period with a FAP greater than 1% was disregarded. During our search for photometric data, we found for 3 variables a mismatch on the cross-reference of the Gaia EDR3 source ID reported by the ASAS-SN photometry database. The stars affected by this mismatch are summarized in Table 4. We investigated the reason for this difference and more details of our analysis are provided for the single stars in section 2.2. It is also noted that for some of the variables (see Table 5) the AAVSO database refers to IBVS 985 data (Maffei 1975) and should be updated to reflect the content of the later update (Maffei and Tosti, 2013).

In general, for each star, bibliographic references available from SIMBAD were reviewed.

Table 4    Incorrect ASAS-SN Gaia EDR3 source ID

| Name | Gaia EDR3 Source ID | ASAS-SN Gaia EDR3 Source ID |
|---|---|---|
| V382 Sct | 4104199043611512192 | 4104199043579502592 |
| V402 Sct | 4104026660866978560 | 4104026660847518976 |
| V406 Sct | 4103139011331374720 | 4103139011331347200 |

Table 5    Variable periods in the AAVSO database vs later paper

| | AAVSO (based on Maffei 1975) | Revised values (Maffei and Tosti, 2013) |
|---|---|---|
| Name | Period (d) | Period (d) |
| V374 Sct | 408.0 | 400 |
| V379 Sct | 413.0 | 417 |
| V383 Sct | 289.0 | 290 |
| V385 Sct | 397.0 | 391 |
| V386 Sct | 426.0 | 444 |
| V387 Sct | 379.0 | 384: |
| V388 Sct | 379.0 | 376 |
| V390 Sct | 319.0 | 322 |
| V392 Sct | 480.0 | 478:: |
| V393 Sct | 381.5 | 378 |
| V394 Sct | 314.0 | 313 |
| V397 Sct | 312.0 | 315 |
| V398 Sct | 368.0 | 366 |
| V399 Sct | 317.0 | 319: |
| V400 Sct | 295.0 | 298 |
| V401 Sct | 474.0 | 477 |
| V403 Sct | 337.0 | 335: |
| V404 Sct | 288.0 | 250: |
| V405 Sct | 307.0 | 362 |
| V406 Sct | 447.0 | 450 |
| V407 Sct | 238.5 | 716 |
| V408 Sct | 321.0 | 318 |
| V409 Sct | 469.0 | 468 |
| V410 Sct | 327.0 | 330 |
| V412 Sct | 408.0 | 416:: |
| V415 Sct | 303.0 | 302 |
| V416 Sct | 300.0 | 301: |
| V417 Sct | 407.0 | 400 |
| V418 Sct | 411.0 | 405 |
| V420 Sct | 261.0 | 264 |
| V422 Sct | 423.0 | 428:: |
| V424 Sct | 474.0 | 468 |
| V425 Sct | 445.0 | 448:: |

## 2.1. Variable stars in the Gaia Colour-Absolute Magnitude Diagram

To get a first general indication of the type of the variable stars listed in Table 1, we first analyzed their position in a Gaia DR2 Colour-Absolute Magnitude Diagram (CAMD).

Figure 1 shows the absolute magnitude $M_G$ vs. color index Bp-Rp for a training sample of about 166,000 variable stars. Different classes of variables are shown in different colors, each one occupying specific areas of the diagram (Jayasinghe et al. 2019). In the same figure, we reported the position of the stars of IBVS 985, as diamond-shaped dots. All photometric data reported in the diagram are not corrected for the interstellar and circumstellar extinction.

Figure 1 highlights that most of the variable stars of IBVS 985 have photometric characteristics compatible with those of a Long Period Variable (LPV) of the training sample, i.e., belonging to the Mira, Semiregular or Irregular variable types.

The red squares indicate the subset of our variables that are classified by WISE (Marton et al. 2016) as candidate Young Stellar Objects (YSO).

Due to the overlap of the various groups of variables in the CAMD, this comparison does not allow a fine classification of the class of variables, but it is still a valid indicator of whether a variable is a candidate to be a Long Period Variable or a YSO.

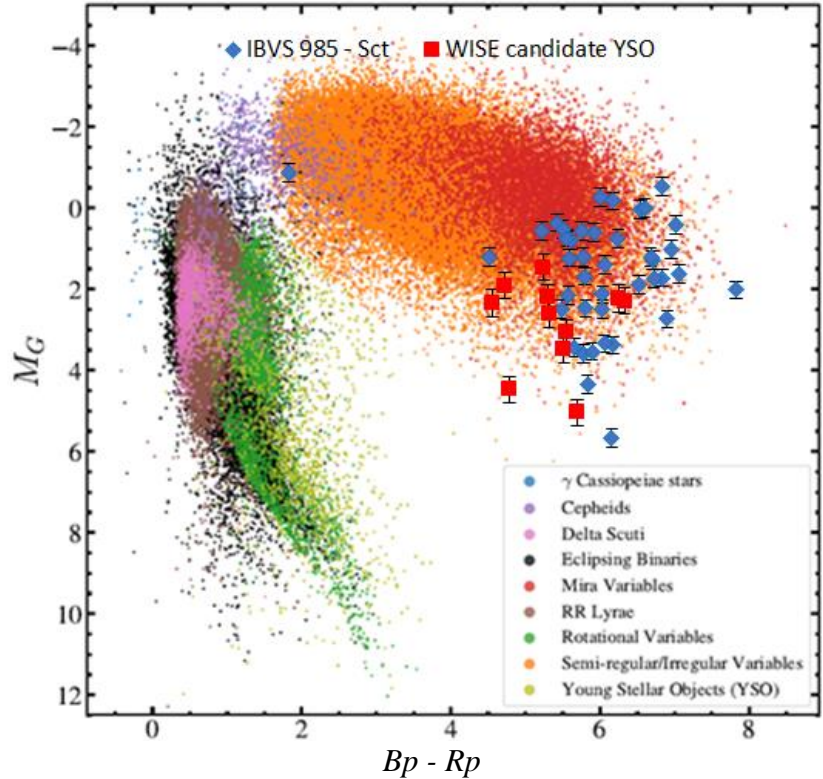

Figure 1 - IBVS 985 variable stars in Scutum vs. Gaia DR2 CAMD

Based on their position in the Gaia DR2 CAMD, we have classified 52 out of 55 variables into the following groups, which are summarized in Table 6:
   a. Not-LPV stars. This group contains only the variable V391 Sct, whose color index Bp-Rp = 1.99 and absolute magnitude $M_G$ = -1.10 are not compatible with a LPV star.

This classification is consistent with the R Coronae Borealis type assigned to this variable by later studies.
  b. Probable YSO. This group contains three stars with a value of absolute magnitude $M_G > 4.3$, that are considered not compatible with a LPV star.
  c. LPV. These 48 stars show photometric characteristics compatible with a Long Period Variable.

For 3 of the 55 stars either the colour index Bp-Rp or the absolute magnitude $M_G$ is not available and therefore they cannot be classified with respect to the CAMD.

Table 6    IBVS 985 variable stars classification based on Gaia DR2 CAMD

| Name | CAMD classification | Name | CAMD classification |
|---|---|---|---|
| NSV 10848 | Not classified | V397 Sct | LPV |
| V375 Sct | Not classified | V398 Sct | LPV |
| V414 Sct | Not classified | V399 Sct | LPV |
| V391 Sct | Not LPV | V400 Sct | LPV |
| NSV 10736 | Probable YSO | V401 Sct | LPV |
| V392 Sct | Probable YSO | V402 Sct | LPV |
| V408 Sct | Probable YSO | V403 Sct | LPV |
| NSV 10832 | LPV | V404 Sct | LPV |
| NSV 10899 | LPV | V405 Sct | LPV |
| V374 Sct | LPV | V406 Sct | LPV |
| V376 Sct | LPV | V407 Sct | LPV |
| V377 Sct | LPV | V409 Sct | LPV |
| V378 Sct | LPV | V410 Sct | LPV |
| V379 Sct | LPV | V411 Sct | LPV |
| V380 Sct | LPV | V412 Sct | LPV |
| V381 Sct | LPV | V413 Sct | LPV |
| V382 Sct | LPV | V415 Sct | LPV |
| V383 Sct | LPV | V416 Sct | LPV |
| V384 Sct | LPV | V417 Sct | LPV |
| V385 Sct | LPV | V418 Sct | LPV |
| V386 Sct | LPV | V419 Sct | LPV |
| V387 Sct | LPV | V420 Sct | LPV |
| V388 Sct | LPV | V421 Sct | LPV |
| V390 Sct | LPV | V422 Sct | LPV |
| V393 Sct | LPV | V423 Sct | LPV |
| V394 Sct | LPV | V424 Sct | LPV |
| V395 Sct | LPV | V425 Sct | LPV |
| V396 Sct | LPV | | |

## 2.2. Detailed data analysis

Details of the relevant information available from the bibliographic references, catalogues and databases and the results of our period analysis and assessment are reported in this section for each star.

**NSV 10736 (V478 Sct)**

The final designation for this variable is V478 Sct (Kazarovets et al. 2006), which is identified with the infrared counterpart 2MASS J18241283-1315552 and Gaia DR2/EDR3 source ID 4152543642155820032. SIMBAD and AAVSO database report this variable as a Mira, in accordance with the original study, which determined a period of 507 days. Based on its Gaia DR2 color index Bp-Rp = 6.148 and absolute magnitude $M_G$ = 5.65 (-1.69; +1.00), we classified this variable as probable YSO.

It is classified as a large amplitude variable in Gaia DR2 (Mowlavi et al. 2021), with an amplitude ΔG = 0.84 mag.

The ASAS-SN Catalog of Variable Stars II (Jayasinghe et al. 2019) identifies this object as source ASASSN-V J182412.84-131555.2 and classifies this variable as a Semiregular, with a mean magnitude V = 17.15, an amplitude of 0.56 mag and a period of 627 days.

We did not perform any analysis of the ASAS-SN photometric data due to insufficient valid measurements available.

**NSV 10832**

This star is identified with the infrared counterpart 2MASS J18274962-1343087 (Nesci, 2018) and Gaia DR2/EDR3 source ID 4104438148705312640. The eclipse nature of this variable is not confirmed by later analyses. The AAVSO database reports this variable as a Semiregular based from the Bochum Galactic Disk Survey II (Hackstein et al. 2015), which refers to this object as source ID GDS J1827496-134308, with a median light curve magnitude r = 16.62, i = 12.78 and an amplitude of 1.65 mag.

This classification is consistent with its Gaia DR2 color index Bp-Rp = 6.515 and absolute magnitude $M_G$ = -0.17 (-1.08; +1.11), that place this variable in the LPV group of Figure 1.

It is classified as a large amplitude variable in Gaia DR2, with an amplitude ΔG = 0.17 mag.

The ASAS-SN Catalog of Variable Stars II does not identify any variable star at equatorial coordinates of NSV 10832. We did not perform any analysis of the ASAS-SN photometric data due to insufficient valid measurements available.

**NSV 10848**

This variable is visible only in two infrared plates (Maffei and Tosti, 1993) and is indicated as a probable Nova, with not a light curve available. A maximum magnitude $I_N$ = 14.4 was observed and the minimum is not defined but fainter than $I_N$ = 18.0 (Maffei and Tosti, 2013).

A later analysis of the plates (Nesci, 2018) determined a brightness I ~ 13.3 with an upper limit B = 17.5 and therefore a very red color index B-I > 4.0. As an alternative to the Nova type, this object could be a cataclysmic variable, which typically undergoes large brightness variations in several years interval. This analysis also proposes as possible visible counterpart the

PanSTARRS/DR1 (Chambers et al. 2016) source ID 91502770485987834 (R.A. 18h 28m 11.67077760s, Dec. -13° 44' 37.2796080", J2000), with magnitudes g = 21.65 ± 0.06, r = 20.05 ± 0.02, i = 19.07 ± 0.02 and z = 18.52 ± 0.02. The source is also present in Gaia DR2 as source ID 4104434785790095104 (G = 19.706, Bp-Rp = 1.626) and in Gaia EDR3 with a different source ID 4104434785801816832 (G = 19.690, Bp-Rp = 2.094). The two Gaia sources are 4 to 6 mas far away from PanSTARRS/DR1 coordinates. Based on its photometric characteristics, the Nova type is the more likely (Nesci 2018).

We noted that the finding chart shown in IBVS 3842 (Maffei and Tosti, 1993) shows a position for this variable significantly different from the one available in the later publication (Maffei and Tosti, 2013), that makes the identification for a candidate counterpart source ID more uncertain (see Figure 2).

Because NSV 10848 is normally a faint object, at PanSTARRS/DR1 coordinates, the ASAS-SN photometry database identifies only the Gaia DR2 source ID 4104434820159940992 (G = 13.869), that is the brighter star about 9" northeast of position of the variable shown in the finding charts. As a consequence, we did not perform any period analysis.

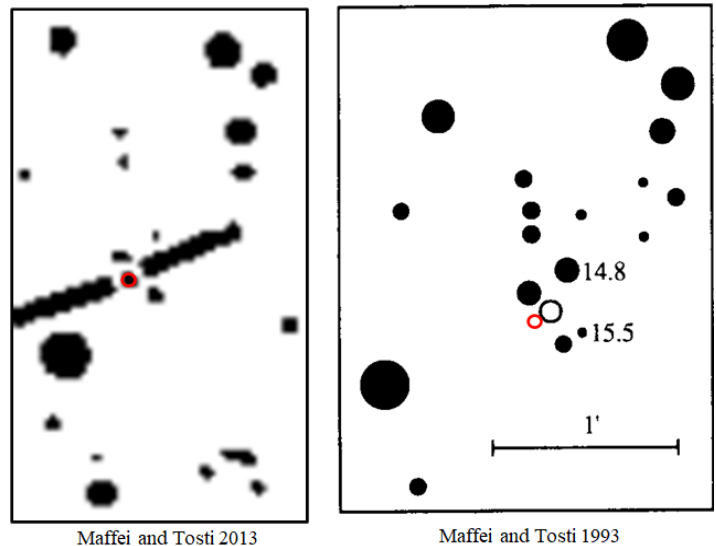

Figure 2 - NSVS 10848 finding charts (the red circles highlight the different positions)

**NSV 10899**

This star is identified with the infrared counterpart 2MASS J18295552-1518384 (Nesci, 2018) and Gaia DR2/EDR3 source ID 4103942582578175744. It is reported as a slow irregular variable by SIMBAD and AAVSO database, in accordance with the original classification. Its Gaia DR2 color index Bp-Rp = 4.771 and absolute magnitude $M_G$ = 1.76 (-1.10; +0.91) are consistent with a LPV star.

It is classified as a large amplitude variable in Gaia DR2, with an amplitude $\Delta G$ = 0.27 mag, and reported as a WISE J182955.52-151838.5 YSO candidate (Marton et al. 2016).

The Bochum Galactic Disk Survey II refers to this variable as source ID GDS J1829555-151838, with a median light curve magnitude r = 16.11, i = 13.94 and an amplitude of 1.60 mag.

No variable star is identified by the ASAS-SN Catalog of Variable Stars II within 10" of the Gaia EDR3 equatorial coordinates of source ID 4103942582578175744.

Our period analysis was performed based on 729 (V) and 1056 (g) valid ASAS-SN observations, applying the Lomb-Scargle and ANOVA methods, and highlighted several potential periods in a range between 3 and 361 days. All periods are statistically significant, but this wide range of potential solutions seems more likely to reflect an irregular behavior of the star's brightness variations, as reported in the original study. Due to this uncertainty, we decided to keep as valid only the solution around 180 days that is confirmed in all filters and methods. With this condition, the weighted average of the four output is 185 ± 8, with mean amplitude $\Delta V$ = 0.056 and $\Delta g$ = 0.044 (see Figure 3).

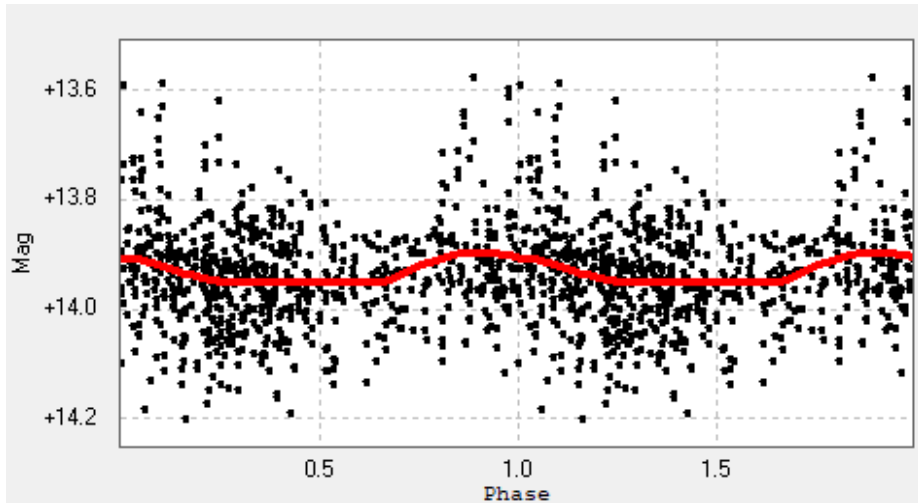

Figure 3 - NSVS 10899 light curve (185 ± 8 days period - weighted average)

**V374 Sct**

This star is identified with the infrared counterpart 2MASS J18230920-1237573 and Gaia DR2/EDR3 source ID 4152798969359304320. The SIMBAD database reports this variable as a Long-Period Variable, in contrast with the original study, which classified it as Eclipse and determined a period of 408 days. This classification is consistent with the Gaia DR2 color index Bp-Rp = 6.919 and absolute magnitude $M_G$ = 0.84 (-0.88; +0.91). It is classified as a large amplitude variable in Gaia DR2, with an amplitude $\Delta G$ = 0.31 mag.

The ASAS-SN Catalog of Variable Stars II refers to this star as ASASSN-V J182309.20-123757.3 and classifies it as a Semiregular, with a mean magnitude V = 13.72, an amplitude of 0.05 mag and a period of 14 days.

Our period analysis was based on 729 and 1083 valid observations available from ASAS-SN in the V and g filter, covering a time span of 1337 and 1081 days, respectively. Applying the Lomb-Scargle method we found a potential period at 29.6 ± 0.3 days, with a curve fit mean amplitude of 0.029 mag, in the g filter (see Figure 4). No reliable solutions were found with the V filter or around 14 and 400 days.

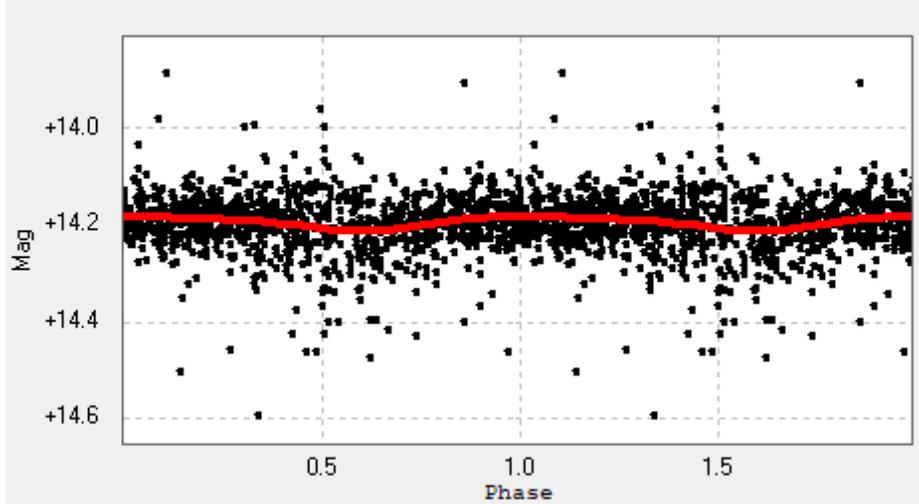

Figure 4 - V374 Sct - g mag vs Phase (29.6 ± 0.3 days period - Lomb-Scargle)

**V375 Sct**
This star is identified with the infrared counterpart 2MASS J18231684-1301356 and Gaia DR2/EDR3 source ID 4152750693951525376. The SIMBAD database reports this variable as a Long-Period Variable, in accordance with the original study, which classified it as Mira and determined a period of 495 days. We could not find its Gaia DR2 color index Bp-Rp, but its Gaia EDR3 color index Bp-Rp = 7.021. The Gaia DR2 absolute magnitude is $M_G$ = 3.68 (-1.43; +1.17). The photometric data are therefore consistent with a LPV star.
It is classified as a large amplitude variable in Gaia DR2, with an amplitude ΔG = 1.00 mag.
No object is identified as a variable star by the ASAS-SN Catalog of Variable Stars II within 10" of the Gaia EDR3 equatorial coordinates of source ID 4152750693951525376.
We did not perform any analysis of the ASAS-SN photometric data due to insufficient valid measurements available.

**V376 Sct**
This star is identified with the infrared counterpart 2MASS J18242192-1251548 and Gaia DR2/EDR3 source ID 4152590749356596864. The SIMBAD database reports this variable as a Mira candidate, in accordance with the original study, which classified it as Mira and determined a period of 590 days.
This classification is consistent with the Gaia DR2 color index Bp-Rp = 6.797 and absolute magnitude $M_G$ = 1.57 (-1.31; +1.26), which place this object in the LPV group of Figure 1.
It is classified as a large amplitude variable in Gaia DR2, with an amplitude ΔG = 0.82 mag.
The ASAS-SN Catalog of Variable Stars II refers to this star as ASASSN-V J182421.92-125154.9 and classifies it as a Semiregular, with a mean magnitude V = 13.58, an amplitude of 0.05 mag and a period of 5 days.
We performed a period analysis, based on 733 and 1086 valid observations available from ASAS-SN in the V and g filters, covering a time span of 1337 and 1080 days, respectively. Applying the Lomb-Scargle method, we could not find any reliable solution for the period in the V filter, but, in the g filter, we found a period of 591 ± 81 days, with a curve fit mean amplitude of 0.02 mag (see Figure 5). We did not identify any reliable solution around 5 days.

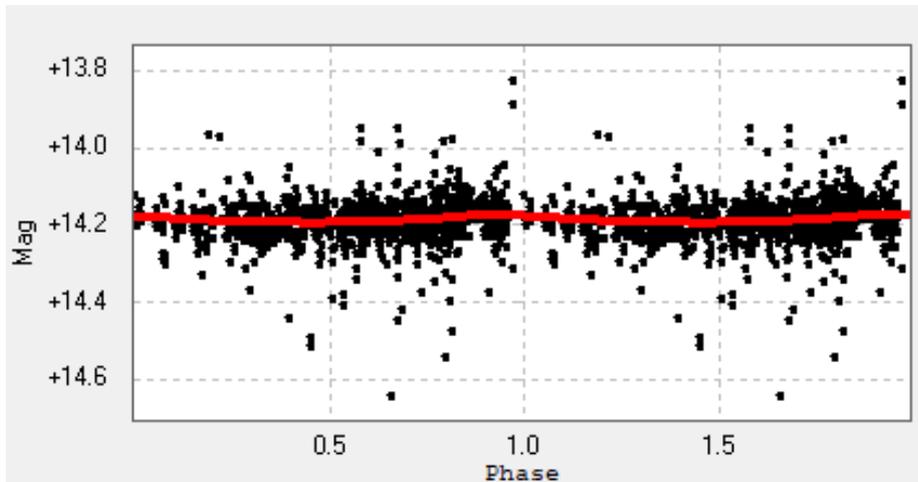

Figure 5 - V376 Sct - g mag vs Phase (591 ± 81 days period - Lomb-Scargle)

**V377 Sct**
This star is identified with the infrared counterpart 2MASS 18245921-1418133 and Gaia DR2/EDR3 source ID 4104298274529689344.
The SIMBAD database reports this variable as a Mira, in accordance with the original study, which determined a period of 522 days. This classification is consistent with the Gaia DR2 color index Bp-Rp = 7.749 and absolute magnitude $M_G$ = 1.88 (-1.31; +1.27), which place this object in the LPV group of Figure 1.
It is classified as a large amplitude variable with an amplitude ΔG = 0.74 mag and a period of 510 ± 134 days (Mowlavi et al. 2018).
The Bochum Galactic Disk Survey II, which refers to this object as source ID GDS J1824592-141813, with a median light curve magnitude i = 14.38 and an amplitude of 1.12 mag.
No object is identified as a variable star by the ASAS-SN Catalog of Variable Stars II within 10" of the Gaia EDR3 equatorial coordinates of source ID 4104298274529689344.
We did not perform any analysis of the ASAS-SN photometric data due to insufficient valid measurements available.

**V378 Sct**
This star is identified with the infrared counterpart 2MASS J18250272-1541063 and Gaia DR2/EDR3 source ID 4097936603350502912.
The SIMBAD database reports this variable as a LPV candidate, in accordance with the original study, which determined a Mira type and a period of 435 days. This classification is consistent with its Gaia DR2 color index Bp-Rp = 6.161 and absolute magnitude $M_G$ = 3.28 (-1.58; +1.23), which place this object in the LPV group of Figure 1. It is classified as a large amplitude variable with an amplitude ΔG = 0.51 mag and a period of 468 ± 61 days.
The Bochum Galactic Disk Survey II, which refers to this object as source ID GDS J1825027-154105, with a median light curve magnitude i = 13.68 and an amplitude of 2.20 mag.

The ASAS-SN Catalog of Variable Stars II classifies this star, ASASSN-V J182502.71-154106.3, as a Semiregular, with a mean magnitude V = 14.76, an amplitude of 0.28 mag and a period of 164 days.

Our period analysis was based on 712 and 1163 valid observations available from ASAS-SN in the V filter and g, covering a time span of 1337 and 1112 days, respectively. Applying the Lomb-Scargle and ANOVA methods we found several solutions in the range 59 ÷ 396 days. Periods around 395 were excluded due to the poor quality of the light curve, which shows a large dispersion of the data and a wide gap in the phase from 0.6 to 0.8 (Figure 6). Two solutions at 60 and 288 days, statistically valid, highlighted in the g filter with the ANOVA method, were disregarded because cannot be found in the analysis of the V measurements. The weighted average of the two valid remaining solutions provides a light curve (Figure 7) with a period of 180 ± 9 days and a curve fit mean amplitude of 0.05÷0.08 mag. We could not find either a reliable epoch for a maximum or a period around or 468 days.

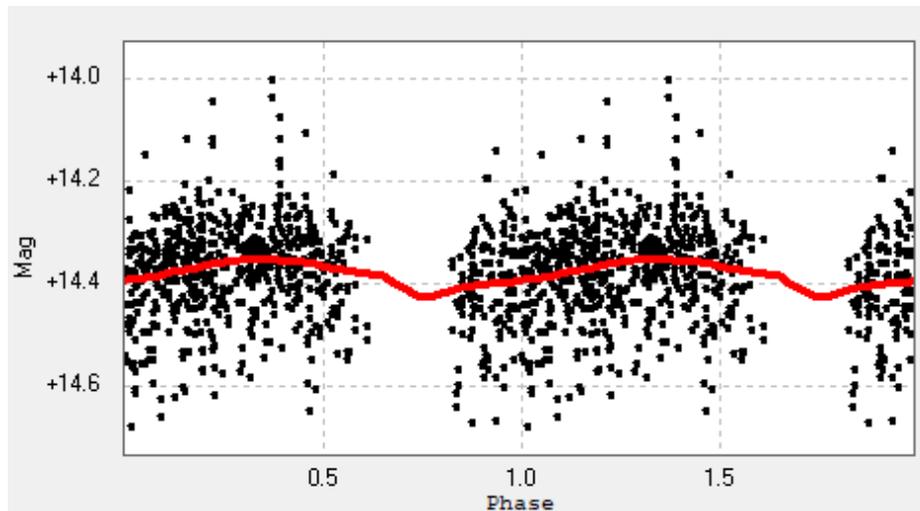

Figure 6 - V378 Sct - V mag vs Phase (396 ± 60 days period - 0.6 ÷ 0.8 gap)

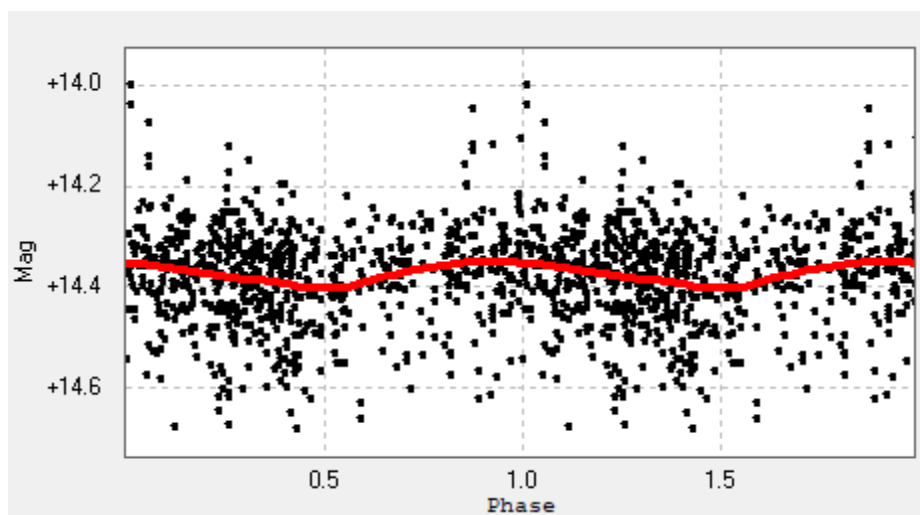

Figure 7 - V378 Sct - V mag vs Phase (180 ± 9 days period - weighted average)

**V379 Sct**

This star is identified with the infrared counterpart 2MASS J18252234-1531597 and Gaia DR2/EDR3 source ID 4097965740393789568.

The SIMBAD database reports this variable as a Mira, in accordance with the original study, which determined a period of 413 days. This classification is consistent with the Gaia DR2 color index Bp-Rp = 6.973 and absolute magnitude $M_G$ = 0.22 (-0.87; +0.88), which place this object in the LPV group of Figure 1.

It is classified as a large amplitude variable in Gaia DR2, with an amplitude $\Delta G$ = 0.47 mag.

The Bochum Galactic Disk Survey II refers to this variable as source ID GDS J1825223-153159, with a median light curve magnitude i = 14.92 and an amplitude of 2.25 mag.

The ASAS-SN Catalog of Variable Stars II refers to this star as ASASSN-V J182522.34-153159.7 and classifies it as a Semiregular, with a mean magnitude V = 15.82, an amplitude of 0.34 mag and a period of 22 days.

We did not perform any analysis of the ASAS-SN photometric data due to insufficient valid measurements available.

**V380 Sct**

This star is identified with the infrared counterpart 2MASS J18254033-1443006 and Gaia DR2/EDR3 source ID 4104201998517789312.

The SIMBAD database reports this variable as a Long-Period Variable, in accordance with the original study, which classified it as Mira and determined a period of 447 days. This classification is consistent with the Gaia DR2 color index Bp-Rp = 5.832 and absolute magnitude $M_G$ = 4.29 (-1.28; +0.84), which place this object in the LPV group of Figure 1.

The Gaia DR2 catalogue of LPV candidates reports a period of 388 ± 54 days for this object, which is also classified as a large amplitude variable in Gaia DR2, with an amplitude $\Delta G$ = 0.49.

The ASAS-SN Catalog of Variable Stars II refers to this star as ASASSN-V J182540.33-144300.7 and classifies it as a Semiregular, with a mean magnitude V = 15.16, an amplitude of 0.16 mag and a period of 13 days.

We performed a period analysis based on 664 and 965 valid observations available from ASAS-SN in the V and g filters, covering a time span of 1330 and 1066 days, respectively. Applying the ANOVA method, we only found, using the observations in the V filter, a period of 29.5 ± 0.2 days with a curve fit mean amplitude of 0.09 mag (Figure 8).

It is noted that a brighter Gaia EDR3 source ID 4104201998549191936 (G = 14.820) is located 4.8 arcsec southwest of V380 Sct. Its Gaia EDR3 color index Bp-Rp = 1.227 makes this second source not compatible with the photometric characteristics of the variable, but we could not exclude that ASAS-SN data and consequently also our solution refer to source ID 4104201998549191936.

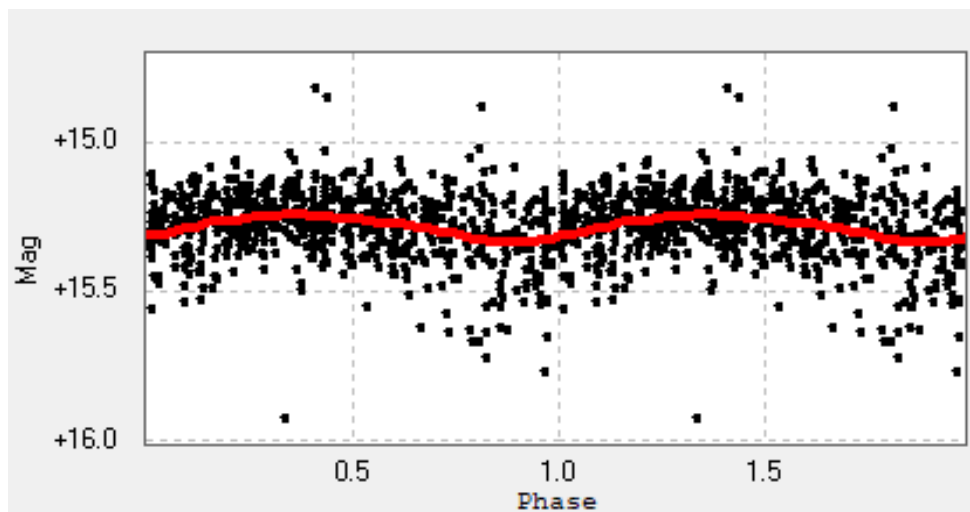

Figure 8 - V380 Sct - V mag vs Phase (29.5 ± 0.2 days period - ANOVA)

**V381 Sct**

This star is identified with the infrared counterpart 2MASS 18255250-1532074 and Gaia DR2/EDR3 source ID 4097962201356698240.

The SIMBAD database reports this variable as a Mira candidate, with an uncertain and unreferenced M spectral type, whilst the original study classified this variable as Semiregular (SRa) and determined a period of 417 days. Both these types are consistent with the Gaia DR2 color index Bp-Rp = 6.681 and absolute magnitude $M_G$ = 1.12 (-1.08; +1.13), which place this object in the LPV group of Figure 1.

It is classified as a large amplitude variable in Gaia DR2, with an amplitude $\Delta G$ = 0.34 mag, and the Bochum Galactic Disk Survey II refers to this variable as source ID GDS J1825525-153207, with a median light curve magnitude i = 14.64 and an amplitude of 0.85 mag.

The ASAS-SN Catalog of Variable Stars II refers to this star as ASASSN-V J182552.51-153207.4 and classifies it as a Semiregular, with a mean magnitude V = 14.14, an amplitude of 0.40 mag and a period of 171 days.

Our period analysis was performed on 717 and 1337 valid observations available from ASAS-SN in the V filter and g, covering a time span of 1337 and 1112 days, respectively. Applying the Lomb-Scargle and ANOVA methods we found several solutions in the range 27 ÷ 394 days. Even if statistically valid, periods greater than 300 days were not considered valid solutions because could not be found with both methods and both filters. The remaining valid solutions were grouped into two distinct groups with weighted average periods 28.5 ± 0.2 (Figure 9) and 183 ± 9 (Figure 10), and a mean curve fit amplitude of 0.09 mag. No reliable solution was found around 417 days.

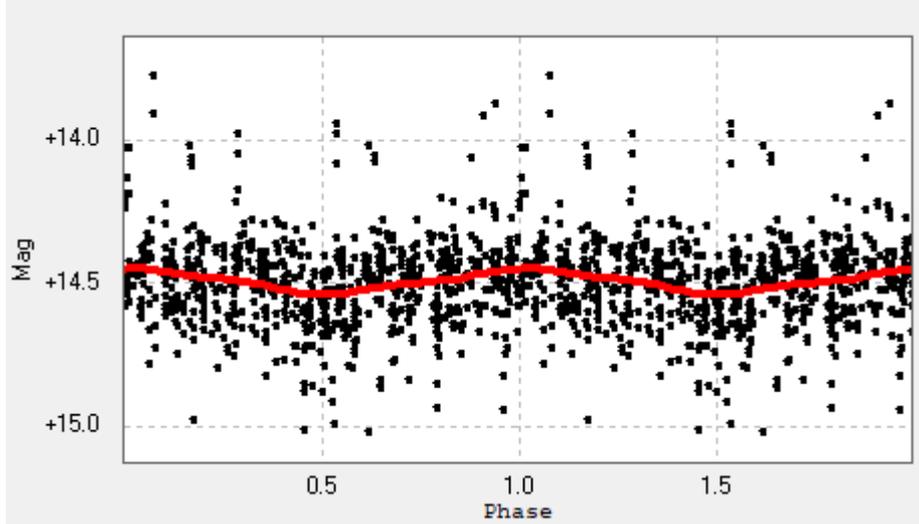

Figure 9 - V381 Sct - V mag vs Phase (28.5 ± 0.2 days period - weighted average)

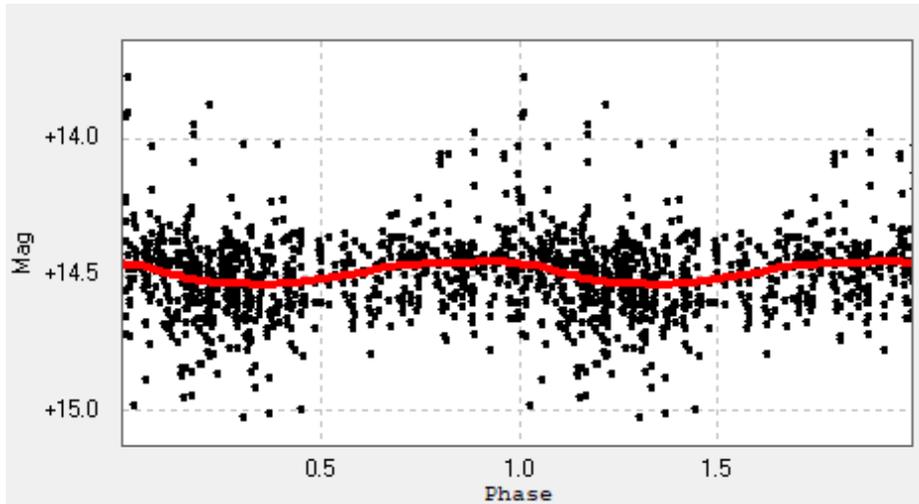

Figure 10 - V381 Sct - V mag vs Phase (183 ± 9 days period - weighted average)

**V382 Sct**

This star is identified with the infrared counterpart 2MASS 18260817-1447151 and Gaia DR2/EDR3 source ID 4104199043611512192. The original study classified it as a possible binary eclipse with an uncertain period of 592 days. The Semiregular (SRc) type classification reported by AAVSO database is based on GCVS catalogue data (November 2019 version), that reflects the preliminary result (Maffei, 1975), and is therefore outdated. The SIMBAD database reports this variable as a LPV star, in accordance with Gaia DR2 LPV candidates catalogue which determined a period of 577 ± 101 days. Also, it is classified as a large amplitude variable in Gaia DR2, with an amplitude ΔG = 0.49 mag.

This classification is consistent with its Gaia DR2 color index Bp-Rp = 6.799 and absolute magnitude $M_G$ = -0.76 (-0.85; +0.88), which place this object in the LPV group of Figure 1.

The ASAS-SN Catalog of Variable Stars II associates V382 Sct to the source ASASSN-V J182607.82-144723.2 and to the different Gaia DR2/EDR3 source ID 4104199043579502592.

This source is classified as a generic, not periodic, variable, with a mean magnitude V = 9.51 and an amplitude of 0.54 mag. The ASAS-SN cross-reference ID is deemed incorrect because source ID 4104199043579502592 is a star 12 arcsec far from 4104199043611512192, with DR2 magnitude G = 17.936 and color index Bp-Rp = 2.181, in contrast with the mean magnitude V = 9.51 of the ASAS-SN light curve. We therefore did not perform any period analysis of ASAS-SN photometric data available for V382 Sct.

**V383 Sct**

This star is identified with the infrared counterpart 2MASS 18263656-1550387 and Gaia DR2/EDR3 source ID 4097895509099999360. The SIMBAD database reports this variable as a Mira, in accordance with the original study, which determined a period of 289 days. This classification is consistent with the Gaia DR2 color index Bp-Rp = 6.860 and absolute magnitude $M_G$ = 2.62 (-1.30; +1.07), which place this object in the LPV group of Figure 1.

It is classified as a large amplitude variable in Gaia DR2, with an amplitude ΔG = 0.67 mag.

At an angular distance of 8" from the Gaia EDR3 equatorial coordinates of ID 4097895509099999360, the ASAS-SN Catalog of Variable Stars II reports source ASASSN-V J182636.57-155038.8 as a Semiregular variable, with a mean magnitude V = 15.65, an amplitude of 0.29 mag and a period of 13 days.

Our period analysis was performed based on 684 valid measurements in the filter V, covering a time span of 1320 days. Applying Lomb-Scargle and ANOVA methods we found a potential period at 13.07 ± 0.03 days (Figure 11). No reliable solution in the g filter and maximum epoch were found.

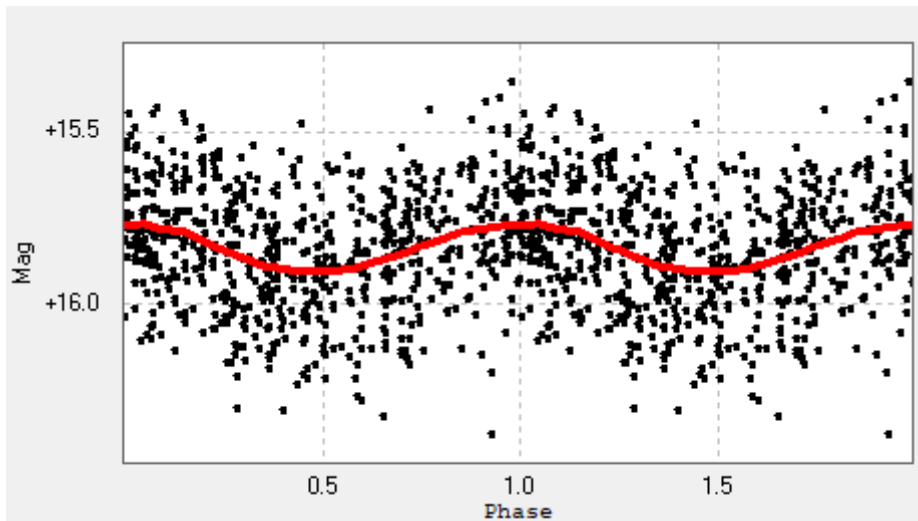

Figure 11 - V383 Sct - V mag vs Phase (13.07 ± 0.03 days period - weighted average)

**V384 Sct**

This star is identified with the infrared counterpart 2MASS J18265495-1540225 and Gaia DR2/EDR3 source ID 4097903824157602560. The SIMBAD database reports this variable as a LPV candidate, in accordance with the uncertain Mira type proposed by the original study, which determined a period of 454 days. This classification is consistent with the Gaia DR2 color index

Bp-Rp = 6.243 and absolute magnitude $M_G$ = 2.09 (-1.03; +1.01), which place this object in the LPV group of Figure 1. It is classified as a large amplitude variable in Gaia DR2, with an amplitude ΔG = 1.04 mag, and reported as a WISE J182654.94-154022.7 YSO candidate. This variable shows also short-time sudden increase of magnitude, which occurs at phase = 0.30, with an amplitude of 0.6 mag (I-N hypersensitized+RG5) and a duration of 12.9 days (Maffei and Tosti, 1995). No object is identified as a variable star by the ASAS-SN Catalog of Variable Stars II within 10" of the Gaia EDR3 equatorial coordinates of source ID 4097903824157602560. We did not perform any analysis of the ASAS-SN photometric data due to insufficient valid measurements available.

**V385 Sct**

This is a bright, S spectral type star (Stephenson, 1994), identified with the infrared counterpart 2MASS J18265916-1441487 and Gaia DR2/EDR3 source ID 4104222614393218816. The original study classified this variable as a Mira and determined a period of 397 days. This result is confirmed by the Gaia DR2 LPV candidate catalogue, which reports a period of 388 ± 22 days, and is in accordance with its Gaia DR2 color index Bp-Rp = 5.445 and absolute magnitude $M_G$ = 0.17 (-1.13; +0.77), which place this variable in the LPV group of Figure 1.

It is classified as a large amplitude variable in Gaia DR2, with an amplitude ΔG = 0.73 mag, and the Bochum Galactic Disk Survey II refers to this variable as source ID GDS J1826591-144148, with a median light curve magnitude r = 10.56, i = 10.74 and an amplitude of 3.75 mag.

The ASAS-SN Catalog of Variable Stars II refers to this star as ASASSN-V J182659.13-144148.6 and classifies it as a Mira variable, with a mean magnitude V = 12.17, an amplitude of 5.03 mag and a period of 404 days.

We performed a period analysis, based on 671 and 1025 valid observations available from ASAS-SN in the V and g filters, covering a time span of 1328 and 1112 days, respectively. Applying the Lomb-Scargle and ANOVA methods to both set of observations we found a weighted average value for the period of 394 ± 3 days (Figure 12). A maximum of the light curve was found at epoch 2458014 ± 2 HJD.

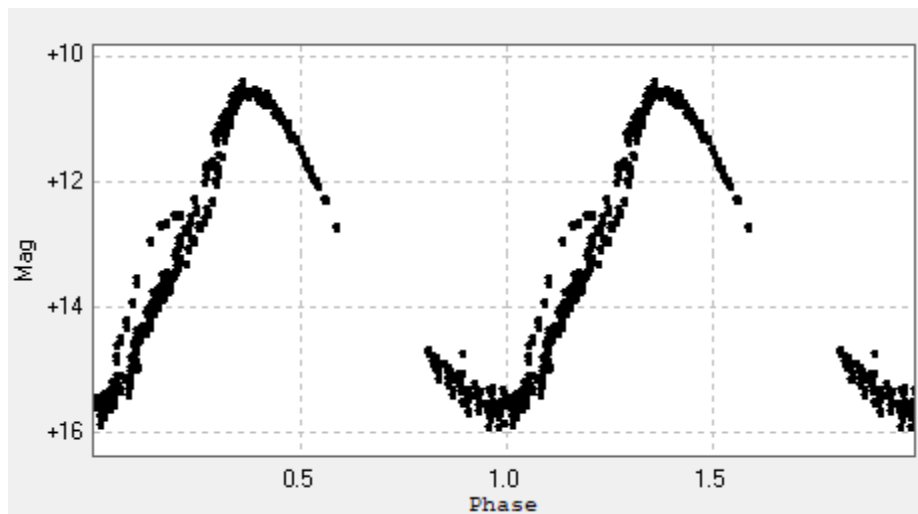

Figure 12 - V385 Sct - V mag vs Phase (394 ± 3 days period - weighted average)

**V386 Sct**

This object is a M8 spectral type star (Stephenson, 1992) and is identified with the counterpart 2MASS 18271563-1549354 and Gaia DR2/EDR3 source ID 4097898532723434368.

The SIMBAD database reports this variable as a Mira, in accordance with the original study, which determined a period of 426 days. This classification is consistent with the Gaia DR2 color index Bp-Rp = 6.690 and absolute magnitude $M_G$ = 1.58 ± 1.07, which place this object in the LPV group of Figure 1.

It is classified as a large amplitude variable in Gaia DR2, with an amplitude ΔG = 0.87 mag.

The Bochum Galactic Disk Survey II refers to this variable as source ID GDS J1827156-154935, with a median light curve magnitude i = 14.86 and an amplitude of 3.51 mag.

This star is reported as an optical variable source in the Integral OMC catalogue (Alfonso-Garzón et al. 2012), with a mean magnitude V = 12.24 ± 0.06, an amplitude of 1.10 mag but with no period defined.

The ASAS-SN Catalog of Variable Stars II refers to this star as ASASSN-V J182715.60-154935.4 and classifies it as a Semiregular, with a mean magnitude V = 14.89, an amplitude of 0.52 mag and a period of 16 days.

We performed a period analysis, based on 954 and 1219 valid observations available from ASAS-SN in the V and g filters, covering a time span of 1335 and 1131 days, respectively. Applying the Lomb-Scargle and ANOVA methods to both set of observations we found a wide range of potential solutions between 41 and 197 days, with a mean fit curve amplitude of 0.12÷0.18 mag. However, we do not consider any of the solutions to be reliable as the values found for the period are not confirmed in both sets of filters and in both methods used for the analysis.

**V387 Sct**

This star is identified with the infrared counterpart 2MASS J18271547-1532243 and Gaia DR2/EDR3 source ID 4097918220888613760.

The SIMBAD database reports this variable as a Mira, in accordance with the classification of the original study, which determined a period of 379 days.

This classification is consistent with the Gaia DR2 color index Bp-Rp = 5.783 and absolute magnitude $M_G$ = 3.50 (-1.39; +0.98), which place this object in the LPV group of Figure 1.

It is classified as a large amplitude variable in Gaia DR2, with an amplitude ΔG = 0.31 mag, and the Bochum Galactic Disk Survey II refers to this variable as source ID GDS J1827154-153224, with a median light curve magnitude i = 15.54 and an amplitude of 3.06 mag.

No object is identified as a variable star by the ASAS-SN Catalog of Variable Stars II within 10" of the Gaia EDR3 equatorial coordinates of source ID 4097918220888613760.

We did not perform any analysis of the ASAS-SN photometric data due to insufficient valid measurements available.

**V388 Sct**

This star is identified with the infrared counterpart 2MASS J18273471-1338110 and Gaia DR2/EDR3 source ID 4104451175395862784. The SIMBAD database reports this variable as a Mira, in accordance with the original study that determined a period of 379 days. This classification is consistent with the Gaia DR2 color index Bp-Rp = 5.624 and absolute magnitude $M_G$ = 0.61 (-0.86; +0.88).

It is classified as a large amplitude variable in Gaia DR2, with an amplitude ΔG = 0.93 mag.

The Bochum Galactic Disk Survey II refers to this variable as source ID GDS J1827347-133810, with a median light curve magnitude r = 15.18, i = 13.63 and an amplitude of 3.08 mag.

No object is identified as a variable star by the ASAS-SN Catalog of Variable Stars II within 10" of the Gaia EDR3 equatorial coordinates of source ID 4104451175395862784.

We performed a period analysis, based on 578 and 860 valid observations available from ASAS-SN in the V filter and g, covering a time span of 1326 and 1085 days, respectively. Applying the Lomb-Scargle and ANOVA methods to the two set of observations in the V and g filters, we found several potential solutions around 200 and 400 days. However, periods around 400 days are not found in all filters and methods and show poor light curve quality, with gap of observations for phases in the range 0.60 ÷ 0.75. We consider as reliable solutions only the periods around 200 day and we found a weighted average value of 196 ± 7 days (Figure 13).

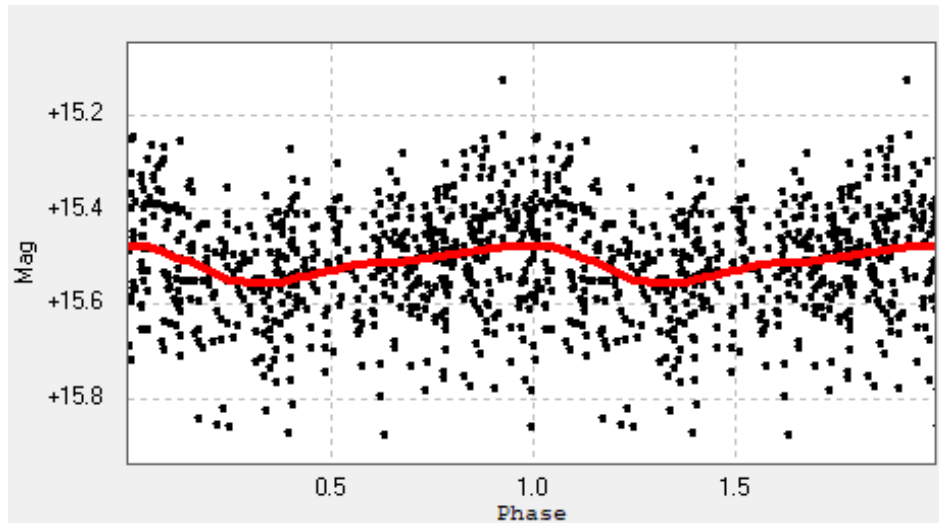

Figure 13 - V388 Sct - V mag vs Phase (196 ± 7 days period - weighted average)

**V390 Sct**

This star is identified with the infrared counterpart 2MASS J18275592-1448415 and Gaia DR2/EDR3 source ID 4104031058914107392.

The SIMBAD database reports this variable as a Mira, in accordance with the classification of the original study, which determined a period of 319 days.

This classification is consistent with the Gaia DR2 color index Bp-Rp = 6.309 and absolute magnitude $M_G$ = 2.15 (-1.20; +1.18), which place this object in the LPV group of Figure 1.

It is classified as a large amplitude variable in Gaia DR2, with an amplitude ΔG = 0.65 mag, and reported as a WISE J182755.94-144841.2 YSO candidate.

The Bochum Galactic Disk Survey II refers to this variable as source ID GDS J1827559-144841, with a median light curve magnitude i = 16.71 and an amplitude of 3.94 mag.

No object is identified as a variable star by the ASAS-SN Catalog of Variable Stars II within 10" of the Gaia EDR3 equatorial coordinates of source ID 4104031058914107392.

We did not perform any analysis of the ASAS-SN photometric data due to insufficient valid measurements available.

**V391 Sct**

This star is identified with the infrared counterpart 2MASS J18280661-1554440, MSX5C G016.1479-02.1803 (Kato, 2001) and Gaia DR2/EDR3 source ID 4097142721628103424. The rapid, irregular variability proposed in the original study is superseded by recent analyses.

This star can be found as a potential U Geminorum in the Catalog of Cataclysmic Variables (Downes et al. 2001). However, this was deemed a misclassification by a later study (Tisserand, 2013), which spectroscopically confirmed the hypothesis that V391 Sct is a R Coronae Borealis. At maximum light, V391 Sct shows pulsations with a period of 124 days (Karambelkar, Viraj 2021).

The R CrB nature of this variable is also confirmed by the ASAS-SN Catalog of Variable Stars II, that refers to this star as ASASSN-V J182806.64-155443.9 and determines a mean magnitude V = 13.44, an amplitude of 0.60 mag and a period of 486 days.

The Bochum Galactic Disk Survey II refers to this variable as source ID GDS J1828066-155444, with a median light curve magnitude r = 12.98, i = 12.26 and an amplitude of 0.33 mag.

Due to the nature of this variable, we did not perform any period analysis of the available ASAS-SN photometric data.

**V392 Sct**

This star is identified with the infrared counterpart 2MASS J18281253-1514385 and Gaia DR2/EDR3 source ID 4103935229609066112.

The SIMBAD database reports this variable as a Mira, in accordance with the original study that determined a period of 480 days. This classification is not consistent with the Gaia DR2 color index Bp-Rp = 5.702 and absolute magnitude $M_G$ = 5.00 (-0.84; +0.61), which place this variable in the YSO group of Figure 1.

It is classified as a large amplitude variable in Gaia DR2, with an amplitude ΔG = 0.65 mag, and reported as a WISE J182812.53-151438.5 YSO candidate.

The Bochum Galactic Disk Survey II refers to this variable as source ID GDS J1828125-151438, with a median light curve magnitude i = 14.24 and an amplitude of 2.56 mag.

The ASAS-SN Catalog of Variable Stars II refers to this star as ASASSN-V J182812.54-151438.5 and classifies it as a Semiregular, with a mean magnitude V = 15.78, an amplitude of 0.71 mag and a period of 42 days.

We did not perform any analysis of the ASAS-SN photometric data due to insufficient valid measurements available.

**V393 Sct**

This is a M7 spectral type star (Skiff, 2014) identified with the infrared counterpart 2MASS J18281626-1412228 and Gaia DR2/EDR3 source ID 4104335795346607872. The SIMBAD database reports this variable as a Mira, in accordance with the original study, which determined a period of 381.5 days. This classification is consistent with the Gaia DR2 color index Bp-Rp =

6.499 and absolute magnitude $M_G$ = 1.75 (-1.53; +1.25), which place this object in the LPV group of Figure 1.

It is classified as a large amplitude variable in Gaia DR2, with an amplitude ΔG = 0.72 mag.

The Bochum Galactic Disk Survey II refers to this variable as source GDS J1828163-141222, with a median light curve magnitude i = 13.47 and an amplitude of 2.02 mag.

The ASAS-SN Catalog of Variable Stars II refers to this star as ASASSN-V J182816.24-141222.6 and classifies it as a Semiregular, with a mean magnitude V = 15.61, an amplitude of 0.69 mag and a period of 388 days.

We performed a period analysis, based on 514 and 739 valid observations available from ASAS-SN in the V filter and g, covering a time span of 1335 and 1066 days, respectively. Applying the Lomb-Scargle and ANOVA methods to the two set of observations in the V and g filters, we found a weighted average for the period of 388 ± 20 days (Figure 14). We found a maximum at epoch 2458596 ± 2 HJD.

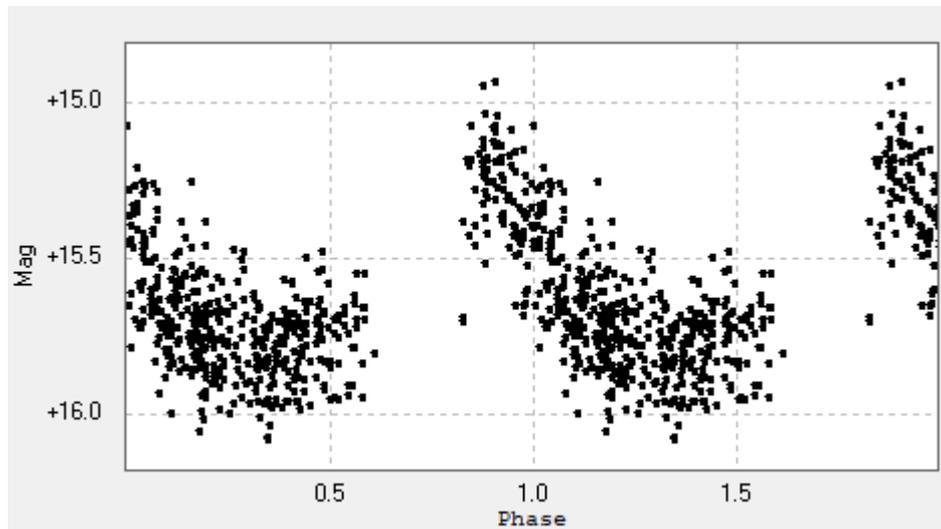

Figure 14 - V393 Sct - V mag vs Phase (388 ± 20 days period - weighted average)

**V394 Sct**

This star is identified with the infrared counterpart 2MASS J18281864-1525508 and Gaia DR2/EDR3 source ID 4103919454096002176.

The SIMBAD database reports this variable as a Mira, in accordance with the original study, that determined a period of 314 days. This classification is consistent with the Gaia DR2 color index Bp-Rp = 5.499 and absolute magnitude $M_G$ = 2.36 (-1.16; +0.86), which place this variable in the LPV group of Figure 1.

It is classified as a large amplitude variable in Gaia DR2, with an amplitude ΔG = 0.84 mag, and the Bochum Galactic Disk Survey II refers to this variable as source ID GDS J1828186-152551, with a median light curve magnitude r = 15.72, i = 13.34 and an amplitude of 1.61 mag.

The ASAS-SN Catalog of Variable Stars II refers to this star as ASASSN-V J182818.64-152550.9 and classifies it as a Semiregular, with a mean magnitude V = 14.61, an amplitude of 0.42 mag and a period of 332 days.

We performed a period analysis, based on 631 and 709 valid observations available from ASAS-SN in the V filter and g, covering a time span of 1328 and 1134 days, respectively. Applying the Lomb-Scargle and ANOVA methods to the two set of observations in the V and g filters, we found a weighted average for the period of 323 ± 18 days (Figure 15). We found a maximum at epoch 2458299 ± 4 HJD.

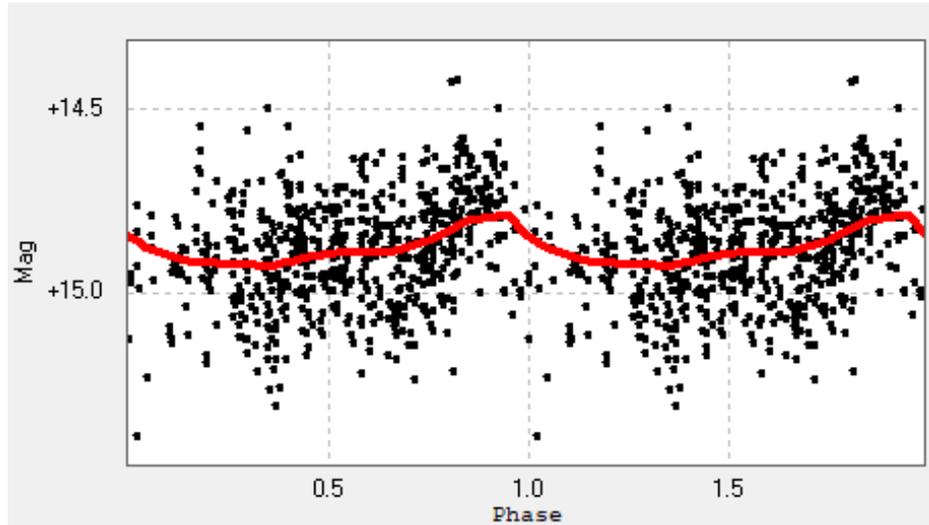

Figure 15 - V394 Sct - V mag vs Phase (323 ± 18 days period - weighted average)

**V395 Sct**

This star is identified with the infrared counterpart 2MASS J18282052-1506009 and Gaia DR2/EDR3 source ID 4103963645044137088.

The SIMBAD database reports this variable as a Mira, in accordance with the original study, that determined a period of 310 days. This classification is consistent with the Gaia DR2 color index Bp-Rp = 5.799 and absolute magnitude $M_G$ = 1.54 (-1.19; +1.10), which place this variable in the LPV group of Figure 1.

The Gaia DR2 classifies this variable as a LPV candidate with a period of 306 ± 10 days and an amplitude of 0.69 mag.

The ASAS-SN Catalog of Variable Stars II refers to this star as ASASSN-V J182820.47-150600.7 and classifies it as a Semiregular, with a mean magnitude V = 13.61, an amplitude of 0.16 mag and a period of 24 days.

We performed a period analysis, based on 729 and 1191 valid observations available from ASAS-SN in the V filter and g, covering a time span of 1337 and 1112 days, respectively. Applying the Lomb-Scargle and ANOVA methods to the set of observations in the V and, we found a weighted average for the period of 314 ± 22 days (Figure 16). No reliable solutions were found for the observations in the filter g.

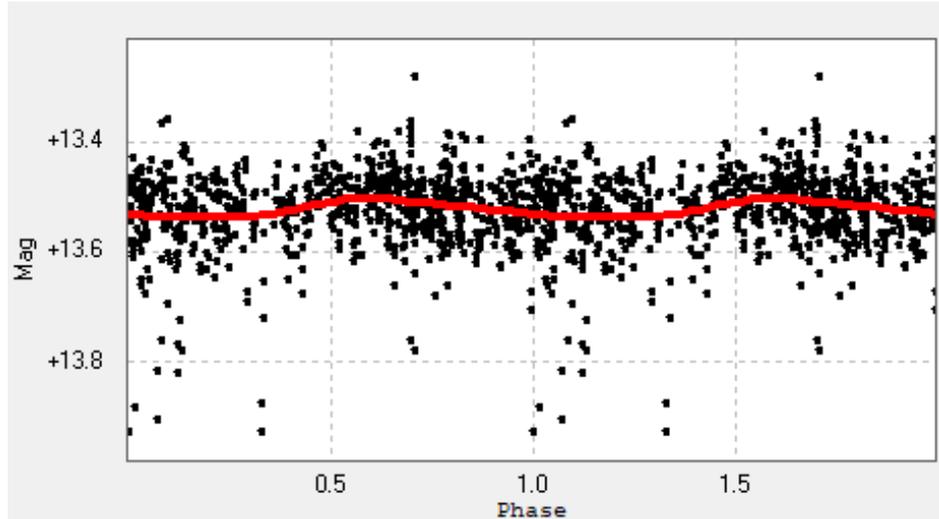

Figure 16 - V395 Sct - V mag vs Phase (314 ± 22 days period - weighted average)

**V396 Sct**

This star is identified with the infrared counterpart 2MASS J18282837-1506279 and Gaia DR2/EDR3 source ID 4103962854825729408.

The SIMBAD database reports this variable as a Mira, in accordance with the original study, that determined a period of 398 days. This classification is consistent with the Gaia DR2 color index Bp-Rp = 5.568 and absolute magnitude $M_G$ = 2.96 ± 1.08, which place this variable in the LPV group of Figure 1.

It is classified as a large amplitude variable in Gaia DR2, with an amplitude $\Delta G$ = 0.36 mag, and reported as a WISE J182828.36-150628.1 YSO candidate.

The ASAS-SN Catalog of Variable Stars II refers to this star as ASASSN-V J182828.37-150627.9 and classifies it as a Semiregular, with a mean magnitude V = 14.09, an amplitude of 0.24 mag and a period of 5 days.

We performed a period analysis, based on 718 and 1177 valid observations available from ASAS-SN in the V filter and g, covering a time span of 1337 and 1112 days, respectively. Applying the Lomb-Scargle and ANOVA methods to the set of observations in the V and g filters we found several potential periods in the range of 250 ÷ 350 days. However, due to the poor quality of the light curves and because the same period cannot be found with both set of observations and methods, we considered these solutions not reliable.

**V397 Sct**

This star is identified with the infrared counterpart 2MASS J18283100-1519244 and Gaia DR2/EDR3 source ID 4103921996717411712. The SIMBAD database reports this variable as a Mira, in accordance with the original study, that determined a period of 312 days. This classification is consistent with the Gaia DR2 color index Bp-Rp = 5.531 and absolute magnitude $M_G$ = 3.38 (-1.25; +0.87), which place this variable in the LPV group of Figure 1.

It is classified as a large amplitude variable in Gaia DR2, with an amplitude $\Delta G$ = 0.62 mag, and reported as a WISE J182830.99-151924.2 YSO candidate.

The ASAS-SN Catalog of Variable Stars II refers to this star as ASASSN-V J182830.99-151923.9 and classifies it as a Semiregular, with a mean magnitude V = 13.39, an amplitude of 0.17 mag and a period of 168 days.

We performed a period analysis, based on 736 and 1195 valid observations available from ASAS-SN in the V filter and g, covering a time span of 1337 and 1112 days, respectively. Applying the Lomb-Scargle and ANOVA methods to the set of observations in the V and g filters we found several potential periods in the range of 110 ÷ 325 days. However, solutions around 110 days were excluded because we could not confirm this period in the analysis with the filter g. The weighted average of the remaining reliable periods is: 306 ± 16 (Figure 17).

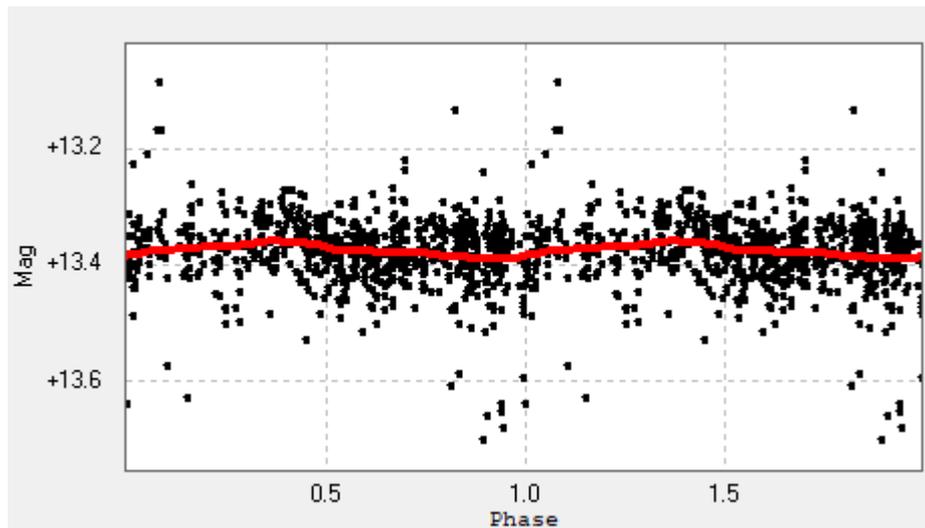

Figure 17 - V397 Sct - V mag vs Phase (306 ± 16 days period - weighted average)

**V398 Sct**

This star is identified with the infrared counterpart 2MASS J18283358-1456333 and Gaia DR2/EDR3 source ID 4104013741603507840. The SIMBAD database reports this variable as a Mira, in accordance with the original study, which determined a period of 368 days. This classification is consistent with the Gaia DR2 color index Bp-Rp = 4.573 and absolute magnitude $M_G$ = 1.05 (-0.97; +0.95), which place this object in the LPV group of Figure 1.

It is classified as a large amplitude variable in Gaia DR2, with an amplitude ΔG = 0.43 mag, and the Bochum Galactic Disk Survey II refers to this variable as source GDS J1828336-145633, with a median light curve magnitude i = 14.12 and an amplitude of 1.58 mag.

No object is identified as a variable star by the ASAS-SN Catalog of Variable Stars II within 10" of the Gaia EDR3 equatorial coordinates of source ID 4104013741603507840.

We did not perform any analysis of the ASAS-SN photometric data due to insufficient valid measurements available.

**V399 Sct**

This star is identified with the infrared counterpart 2MASS J18283523-1456186 and Gaia DR2/EDR3 source ID 4104013771581820416. The SIMBAD database reports this variable as a Mira, in accordance with the original study, which determined a period of 317 days. This classification is consistent with the Gaia DR2 color index Bp-Rp = 5.593 and absolute magnitude $M_G$ = 2.02 (-1.27; +1.23), which place this object in the LPV group of Figure 1.

It is classified as a large amplitude variable in Gaia DR2, with an amplitude ΔG = 0.49 mag.

No object is identified as a variable star by the ASAS-SN Catalog of Variable Stars II within 10" of the Gaia EDR3 equatorial coordinates of source ID 4104013771581820416.

We did not perform any analysis of the ASAS-SN photometric data due to insufficient valid measurements available.

**V400 Sct**

This star is identified with the infrared counterpart 2MASS J18284236-1522525 and Gaia DR2/EDR3 source ID 4103919351017148160. The SIMBAD database reports this variable as a Mira, in accordance with the original study, which determined a period of 295 days. This classification is consistent with the Gaia DR2 color index Bp-Rp = 5.775 and absolute magnitude $M_G$ = 1.04 (-1.23; +0.94), which place this object in the LPV group of Figure 1.

It is classified as a large amplitude variable in Gaia DR2, with an amplitude ΔG = 0.28 mag.

The ASAS-SN Catalog of Variable Stars II refers to this star as ASASSN-V J182842.35-152252.4 and classifies it as a Semiregular, with a mean magnitude V = 15.53, an amplitude of 0.45 mag and a period of 297 days.

We performed a period analysis, based on 261 and 839 valid observations available from ASAS-SN in the V filter and g, covering a time span of 1296 and 1135 days, respectively. We found two potential solutions that are statistically significant applying Lomb-Scargle and ANOVA methods to both set of observations in the V and g filters. The weighted average values of the solutions shown in Figures 18 and 19 are: 113 ± 2 and 318 ± 13 days.

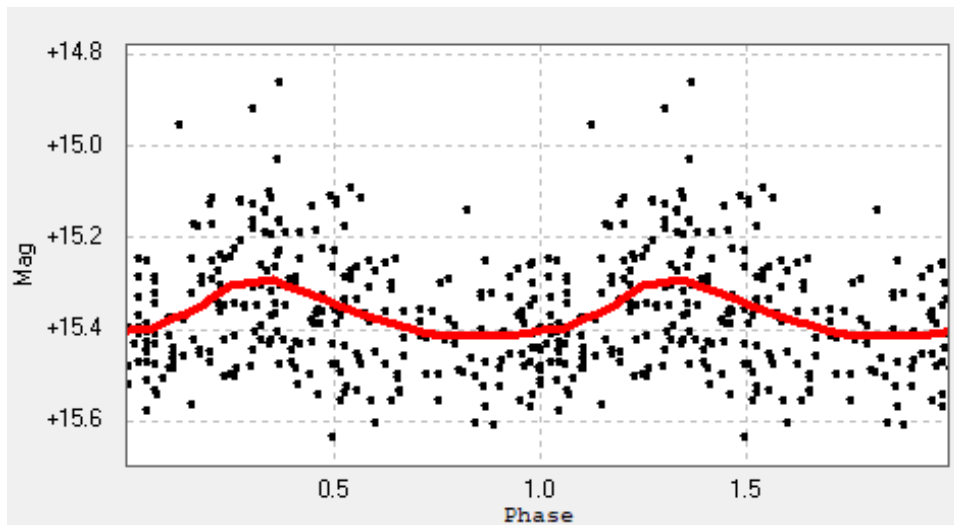

Figure 18 - V400 Sct - V mag vs Phase (113 ± 2 days period - weighted average)

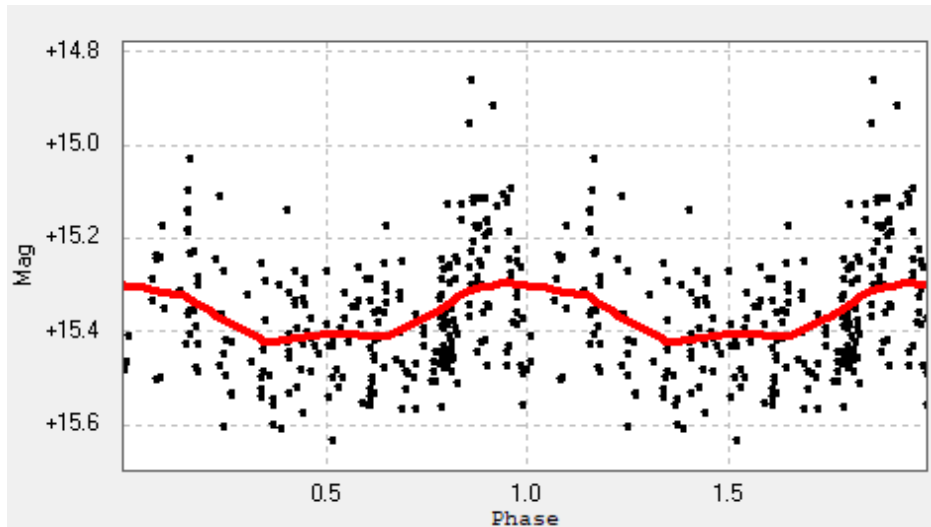

Figure 19 - V400 Sct - V mag vs Phase (318 ± 13 days period - weighted average)

**V401 Sct**

This star is identified with the infrared counterpart 2MASS J18285417-1429201 and Gaia DR2/EDR3 source ID 4104137432283390464. The SIMBAD database reports this variable as a Mira, in accordance with the original study, which determined a period of 474 days. This classification is consistent with the Gaia DR2 color index Bp-Rp = 6.648 and absolute magnitude $M_G$ = 1.04 (-1.24; +1.05), which place this object in the LPV group of Figure 1.

The Gaia DR2 classifies this variable as a LPV candidate with a period of 432 ± 92 days and an amplitude of 0.40 mag.

The Bochum Galactic Disk Survey II refers to this variable as source GDS J1828542-142920, with a median light curve magnitude i = 15.17 and an amplitude of 1.59 mag.

No object is identified as a variable star by the ASAS-SN Catalog of Variable Stars II within 10" of the Gaia EDR3 equatorial coordinates of source ID 4104137432283390464.

We did not perform any analysis of the ASAS-SN photometric data due to insufficient valid measurements available.

**V402 Sct**

This star is identified with the infrared counterpart 2MASS J18290073-1443550 and Gaia DR2/EDR3 source ID 4104026660866978560. The SIMBAD database reports this variable as a Mira, in accordance with the original study, which determined a period of 312 days. This classification is consistent with the Gaia DR2 color index Bp-Rp = 5.246 and absolute magnitude $M_G$ = 0.37 (-1.06; +1.08), which place this object in the LPV group of Figure 1.

It is classified as a large amplitude variable in Gaia DR2, with an amplitude $\Delta G$ = 0.90 mag, and the Bochum Galactic Disk Survey II refers to this variable as source GDS J1829007-144354, with a median light curve magnitude r = 13.92, i = 14.21 and an amplitude of 3.61 mag.

The ASAS-SN Catalog of Variable Stars II refers to this star, ASASSN-V J182900.67-144354.8, with a different Gaia EDR3 source ID 4104026660847518976, and classifies it as a Semiregular, with a mean magnitude V = 14.55, an amplitude of 1.32 mag and a period of 297 days. Source

ID 4104026660847518976 is a fainter object (G = 16.514), with no color index defined, and is very close (0.7 arcsec) to source ID 4104026660866978560. Based on the original infrared range measurements of V402 Sct, the ASAS-SN cross reference ID is probably incorrect because the mean magnitude G of the variable is expected to be brighter.

We performed a period analysis, based on 638 and 1100 valid observations available from ASAS-SN in the V filter and g, covering a time span of 1326 and 1112 days, respectively. Applying Lomb-Scargle and ANOVA methods to both set of observations in the V and g filters we found four statistically valid solutions and a weighted average value for the period of 314 ± 6 days (Figure 20). We also identified an epoch for the maximum of the light curve at 2459041 ± 2 HJD.

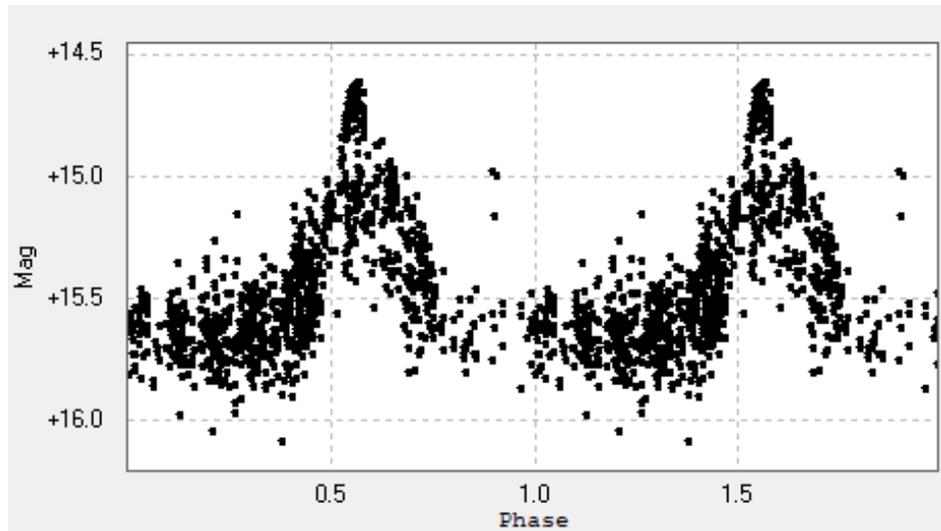

Figure 20 - V402 Sct - V mag vs Phase (314 ± 6 days period - weighted average)

**V403 Sct**

This star is identified with the infrared counterpart 2MASS J18290273-1446582 and Gaia DR2/EDR3 source ID 4104023018734396800. The SIMBAD database reports this variable as a Mira, in accordance with the original study, which determined a period of 337 days. This classification is consistent with the Gaia DR2 color index Bp-Rp = 6.061 and absolute magnitude $M_G$ = 1.24 (-1.02; +1.06), which place this object in the LPV group of Figure 1.

It is classified as a large amplitude variable in Gaia DR2, with an amplitude ΔG = 0.57 mag.

No object is identified as a variable star by the ASAS-SN Catalog of Variable Stars II within 10" of the Gaia EDR3 equatorial coordinates of source ID 4104023018734396800.

We did not perform any analysis of the ASAS-SN photometric data due to insufficient valid measurements available.

**V404 Sct**

This star is identified with the infrared counterpart 2MASS J18291283-1537377 and Gaia DR2/EDR3 source ID 4103161894951509504. The SIMBAD database reports this variable as a Mira, in accordance with the original study, which determined a period of 288 days. This classification is consistent with the Gaia DR2 color index Bp-Rp = 4.607 and absolute magnitude $M_G$ = 2.20 (-0.91; +0.90), which place this object in the LPV group of Figure 1.

It is classified as a large amplitude variable in Gaia DR2, with an amplitude ΔG = 0.59 mag, and the Bochum Galactic Disk Survey II refers to this variable as source GDS J1829128-153737, with a median light curve magnitude i = 15.02 and an amplitude of 1.59 mag.

The ASAS-SN Catalog of Variable Stars II classifies this star, ASASSN-V J182912.84-153737.7, as a Semiregular, with a mean magnitude V = 15.57, an amplitude of 0.34 mag and a period of 18 days. We did not perform any analysis of the ASAS-SN photometric data due to insufficient valid measurements available.

**V405 Sct**

This star is identified with the infrared counterpart 2MASS 18292248-1507598 and Gaia DR2/EDR3 source ID 4103948939130077056. The SIMBAD database reports this variable as a Mira, in accordance with the original study, which determined a period of 307 days. This classification is consistent with the Gaia DR2 color index Bp-Rp = 5.337 and absolute magnitude $M_G$ = 2.48 (-1.11; +1.10), which place this object in the LPV group of Figure 1.

It is classified as a large amplitude variable in Gaia DR2, with an amplitude ΔG = 0.74 mag, and reported as a WISE J182922.48-150759.6 YSO candidate. In the Bochum Galactic Disk Survey II this star is associated to source ID GDS J1829225-150800, with a median light curve magnitude i = 13.39 and an amplitude of 0.69 mag.

The ASAS-SN Catalog of Variable Stars II refers to this star as ASASSN-V J182922.49-150759.8 and classifies it as a Semiregular, with a mean magnitude V = 15.28, an amplitude of 0.25 mag and a period of 42 days.

We performed a period analysis, based on 501 and 850 valid observations available from ASAS-SN in the V filter and g, covering a time span of 1326 and 1074 days, respectively. Applying the Lomb-Scargle and ANOVA methods, we could not find any reliable solution for the period.

**V406 Sct**

This star is identified with the infrared counterpart 2MASS J18292375-1547258 and Gaia DR2/EDR3 source ID 4103139011331374720. The original study determined for this variable an uncertain Mira type with a period of 447 days. This classification is consistent with the Gaia DR2 color index Bp-Rp = 6.055 and absolute magnitude $M_G$ = 3.24 (-0.67; +0.52), which place this object in the LPV group of Figure 1.

The Gaia DR2 classifies this variable as a LPV candidate with a period of 459 ± 31 days and an amplitude of 0.58 mag.

At a distance of 6.6 arcsec from Gaia DR2 4103139011331374720, the ASAS-SN Catalog of Variable Stars II identifies the source ASASSN-V J182923.77-154732.1 as V406 Sct, but with a different Gaia EDR3 ID 4103139011331347200. The ASAS-SN cross-reference ID is deemed incorrect because the JHK magnitudes reported in the database are those of source ID 4103139011331374720. Moreover, this object has a magnitude G = 19.045 that is not compatible with the V mean magnitude measured by ASAS-SN.

We performed a period analysis, based on 767 and 1214 valid observations available from ASAS-SN in the V filter and g, covering a time span of 1337 and 1112 days, respectively. Applying the Lomb-Scargle and ANOVA methods, we highlighted several potential periods, that confirm the uncertainty on the type and on the periodicity for this variable. None of the solutions was found with all filters and methods. The weighted average values for the periods we found are: 192 ± 16, 291 ± 29 and 455 ± 66 days. The light curve with the longer period is shown in Figure 21. A maximum was also identified at epoch 2457941 ± 10 HJD.

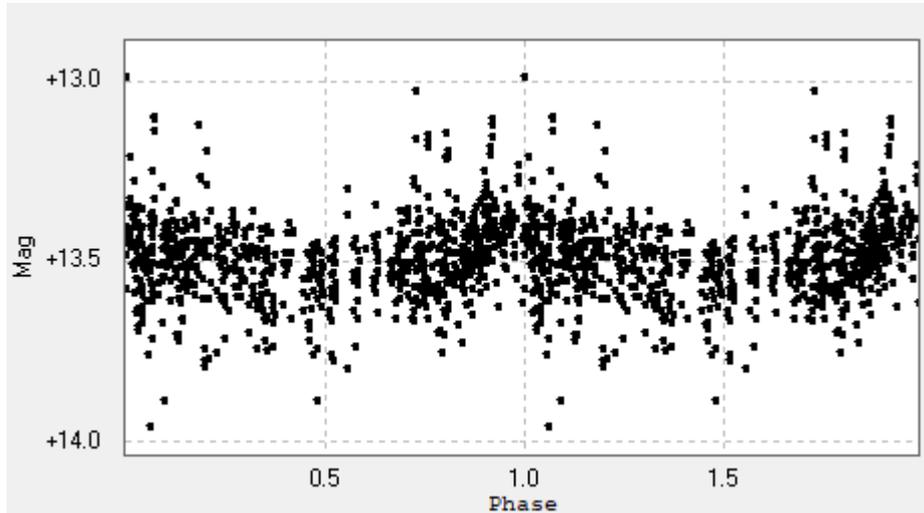

Figure 21 - V406 Sct - V mag vs Phase (455 ± 66 days period - weighted average)

**V407 Sct**

This star is identified with the infrared counterpart 2MASS J18293219-1548396 and Gaia DR2/EDR3 source ID 4103138148004853248. The original study determined for this variable a Mira type with a period of 238.5 days. This classification is consistent with the Gaia DR2 color index Bp-Rp = 5.996 and absolute magnitude $M_G$ = -0.47 (-0.94; +0.88), which place this object in the LPV group of Figure 1.

The Gaia DR2 classifies this variable as a LPV candidate with a period of 231 ± 5 days and an amplitude of 0.36 mag.

The Bochum Galactic Disk Survey II refers to this variable as source GDS J1829322-154839, with a median light curve magnitude r = 12.52, i = 10.70 and an amplitude of 3.63 mag.

The ASAS-SN Catalog of Variable Stars II classifies this star, ASASSN-V J182932.15-154839.1, as a Mira, with a mean magnitude V = 14.44, an amplitude of 1.80 mag and a period of 178 days.

We performed a period analysis, based on 584 and 714 valid observations available from ASAS-SN in the V filter and g, covering a time span of 1325 and 1114 days, respectively.

Applying the Lomb-Scargle and ANOVA methods to both set of observations, we highlighted a potential solution with a weighted average period of 234 ± 1 days (Figure 22). A statistically valid solution was also found for a weighted average period of 118 ± 2 days but was excluded due to the poor quality of the light curve. A maximum was identified at epoch 2459026 ± 4 HJD.

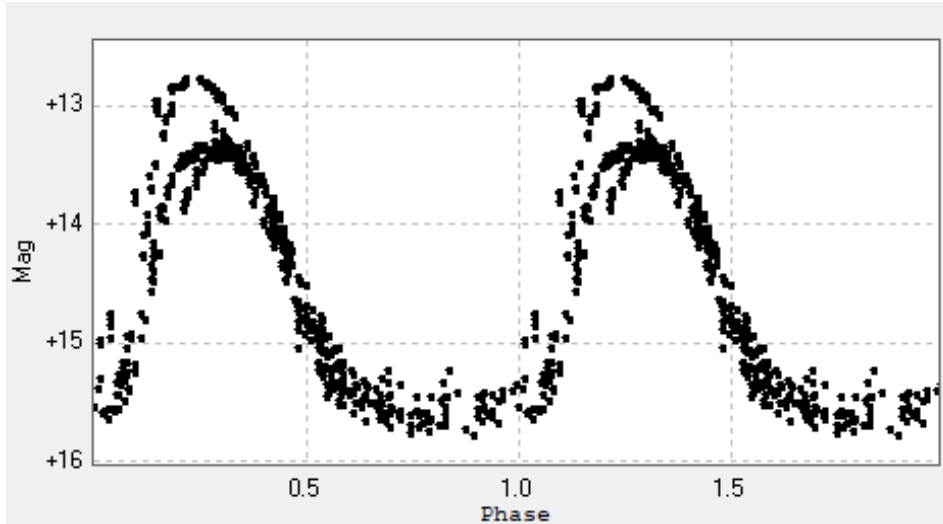

Figure 22 - V407 Sct - V mag vs Phase (234 ± 1 days period - weighted average)

**V408 Sct**

This star is identified with the infrared counterpart 2MASS J18293898-1446124 and Gaia DR2/EDR3 source ID 4104069468721467136.

The original study determined for this variable a Mira type with a period of 321 days. This classification is not consistent with the Gaia DR2 color index Bp-Rp = 4.832 and absolute magnitude $M_G$ = 4.41 (-1.75; +1.21), which place this object in the YSO group of Figure 1.

It is classified as a large amplitude variable in Gaia DR2, with an amplitude ΔG = 0.52 mag, and reported as a WISE J182938.97-144612.4 YSO candidate. In the Bochum Galactic Disk Survey II this star is associated to source ID GDS J1829389-144612, with a median light curve magnitude i = 15.11 and an amplitude of 2.86 mag.

The ASAS-SN Catalog of Variable Stars II refers to this star as ASASSN-V J182938.98-144612.4 and classifies it as a Semiregular, with a mean magnitude V = 14.82, an amplitude of 0.19 mag and a period of 166 days.

We performed a period analysis, based on 675 and 1093 valid observations available from ASAS-SN in the V filter and g, covering a time span of 1335 and 1110 days, respectively.

Applying the Lomb-Scargle and ANOVA methods to both set of observations, we highlighted a potential solution with a weighted average period of 29.4 ± 0.2 days (Figure 23). A valid solution was also highlighted at 313 ± 22 days but was excluded because is not found in the analyses with the filter g. A maximum was identified at epoch 2458302 ± 2 HJD.

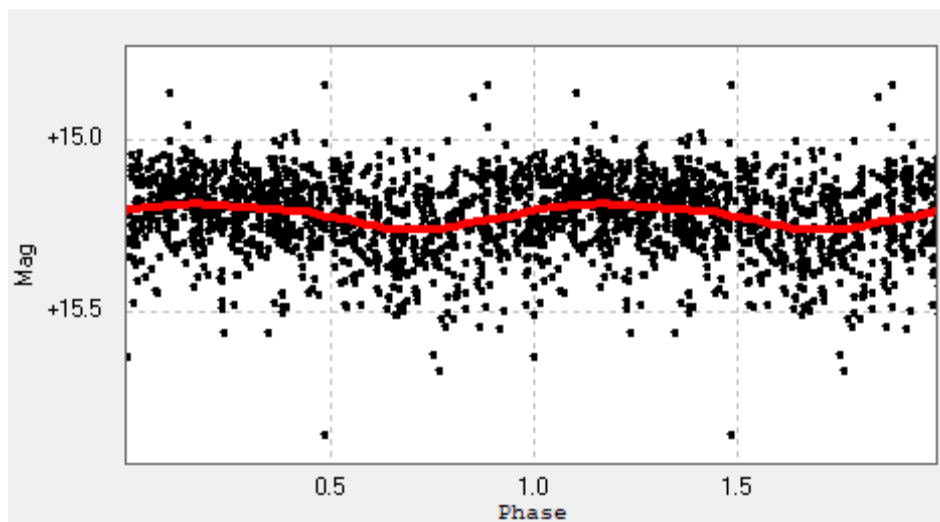

Figure 23 - V408 Sct - V mag vs Phase (29.4 ± 0.2 days period - weighted average)

**V409 Sct**

This star is identified with the infrared counterpart 2MASS 18294001-1400178 (Nesci, 2018) and Gaia DR2/EDR3 source ID 4104366341161294976. The SIMBAD database reports erroneously this variable as source 2MASS J18293968-1359367 and Gaia DR2/EDR3 4104366341197354624, because of original coordinates offset of 30 arcsec. The original study classified V409 Sct as a Mira with a period of 469 days. This classification is consistent with the Gaia DR2 color index Bp-Rp = 5.669 and absolute magnitude $M_G$ = 3.35 (-1.66; +1.22), which place this object in the LPV group of Figure 1.

It is classified as a large amplitude variable in Gaia DR2, with an amplitude $\Delta G$ = 1.07 mag.

The ASAS-SN Catalog of Variable Stars II refers to this star as ASASSN-V J182940.01-140017.8 and classifies it as a Semiregular, with a mean magnitude V = 15.12, an amplitude of 0.26 mag and a period of 342 days.

We performed a period analysis, based on 630 and 1118 valid observations available from ASAS-SN in the V filter and g, covering a time span of 1326 and 1151 days, respectively.

We highlighted two potential solutions at 27.6 ± 0.3 and 214 ± 21 days with the ANOVA method for the observations in the filter g. However, we consider these solutions not reliable, because cannot be found with Lomb-Scargle method and/or the set of observations in the V filter.

**V410 Sct**

This star is identified with the infrared counterpart 2MASS J18295301-1457534 and Gaia DR2/EDR3 source ID 4103969348817568896. The SIMBAD database reports this variable as a Mira, in accordance with the original study, which determined a period of 327 days. This classification is consistent with the Gaia DR2 color index Bp-Rp = 5.607 and absolute magnitude $M_G$ = 1.06 (-0.93; +0.96), which place this object in the LPV group of Figure 1.

It is classified as a large amplitude variable in Gaia DR2, with an amplitude $\Delta G$ = 0.51 mag, and the Bochum Galactic Disk Survey II refers to this variable as source GDS J1829533-145754, with a median light curve magnitude i = 15.39 and an amplitude of 1.88 mag.

No object is identified as a variable star by the ASAS-SN Catalog of Variable Stars II within 10" of the Gaia EDR3 equatorial coordinates of source ID 4103969348817568896.

We did not perform any analysis of the ASAS-SN photometric data due to insufficient valid measurements available.

**V411 Sct**

This star is identified with the infrared counterpart 2MASS J18275904-1343220 and Gaia DR2/EDR3 source ID 4104437912536530432. The SIMBAD database reports this variable as a LPV Candidate, in accordance with the original study, which classified it as Semiregular and determined a period of 457 days. Its Gaia DR2 color index Bp-Rp = 6.569 and absolute magnitude $M_G$ = -0.20 (-0.89; +0.83) are consistent with a LPV star of Figure 1.

It is classified as a large amplitude variable in Gaia DR2, with an amplitude $\Delta G$ = 0.49 mag.

The Bochum Galactic Disk Survey II refers to this variable as source ID GDS J1827595-134321, with a median light curve magnitude r = 15.91, i = 12.97 and an amplitude of 1.52 mag.

This variable shows also short-time sudden increase of magnitude, which occurs at phase = 0.75, with an amplitude of 0.5 mag (I-N hypersensitized+RG5) and a duration less than 24.0 days (Maffei and Tosti, 1995).

No object is identified as a variable star by the ASAS-SN Catalog of Variable Stars II within 10" of the Gaia EDR3 equatorial coordinates of source ID 4104437912536530432.

We did not perform any analysis of the ASAS-SN photometric data due to insufficient valid measurements available.

**V412 Sct**

This star is identified with the infrared counterpart 2MASS J18295881-1410102 and Gaia DR2/EDR3 source ID 4104174304682462976. The SIMBAD database reports this variable as a Mira, in accordance with original study, which determined a period of 408 days. Its Gaia DR2 color index Bp-Rp = 6.030 and absolute magnitude $M_G$ = 1.99 (-1.16; +1.17) are consistent with a LPV star of Figure 1.

It is classified as a large amplitude variable in Gaia DR2, with an amplitude $\Delta G$ = 0.58 mag.

The Bochum Galactic Disk Survey II refers to this variable as source ID GDS J1829587-141010, with a median light curve magnitude i = 15.23 and an amplitude of 2.97 mag.

No object is identified as a variable star by the ASAS-SN Catalog of Variable Stars II within 10" of the Gaia EDR3 equatorial coordinates of source ID 4104174304682462976.

We did not perform any analysis of the ASAS-SN photometric data due to insufficient valid measurements available.

**V413 Sct**

This star is identified with the infrared counterpart 2MASS J18300233-1528300 and Gaia DR2/EDR3 source ID 4103185397013474432. The SIMBAD database reports this variable as a LPV Candidate, in accordance with original study, which classified it as a Mira and determined a period of 333 days. Its Gaia DR2 color index Bp-Rp = 6.029 and absolute magnitude $M_G$ = 2.36 (-1.16; +1.18) are consistent with a LPV star of Figure 1.

It is classified as a large amplitude variable in Gaia DR2, with an amplitude ΔG = 0.49 mag.
The ASAS-SN Catalog of Variable Stars II refers to this star as ASASSN-V J183002.33-152830.2 and classifies it as a slow Irregular variable, with a mean magnitude V = 15.52 and an amplitude of 0.63 mag.
We performed a period analysis, based on 537 and 884 valid observations available from ASAS-SN in the V filter and g, covering a time span of 1330 and 1112 days, respectively but we could not identify a solution valid in both filters and methods. Our analysis seems to confirm the irregularity of brightness variation of this star.

**V414 Sct**
This star is identified with the infrared counterpart 2MASS J18301384-1527224 and Gaia DR2/EDR3 source ID 4103185637531623552. The original study classified this variable as a Mira and determined a period of 529 days. This classification is consistent with its Gaia EDR3 color index Bp-Rp = 5.880 and absolute magnitude $M_G$ = 2.59 (-0.81; +0.77). The Gaia DR2 color index Bp-Rp is not available for V414 Sct.
It is noted that within a radius of 2 arcsec there is another, brighter, Gaia DR2/EDR3 source (4103185633161346560), with a similar magnitude (G = 15.482, EDR3) but a bluer color index (Bp-Rp = 1.244, EDR3).
It is classified as a large amplitude variable in Gaia DR2, with an amplitude ΔG = 0.84 mag, and is reported as a WISE J183013.85-152722.4 YSO candidate.
The ASAS-SN Catalog of Variable Stars II classifies this star, ASASSN-V J183013.85-152722.4, as a generic variable, with a mean magnitude V = 15.62 and an amplitude of 0.36 mag.
We performed a period analysis, based on 263 and 481 valid observations available from ASAS-SN in the V filter an g, covering a time span of 1320 and 1065 days, respectively. Applying the ANOVA and Lomb-Scargle methods, we could not find any reliable solution for the period in the V filter. With the set of observations in the g filter we could highlight two potential solutions with period 308 ± 32 days (weighted average, Figure 24) and 592 ± 77 days (ANOVA method, Figure 25). We also identified a maximum at epoch 2458952 ± 3 HJD.

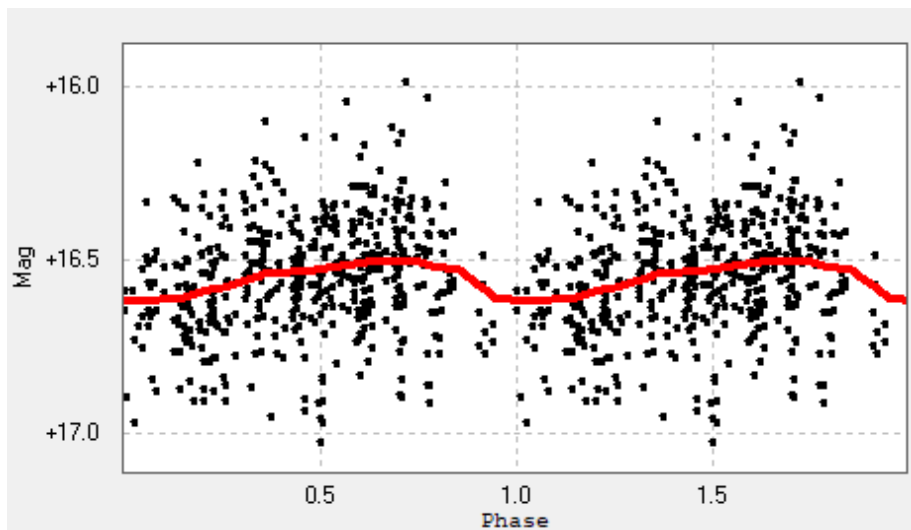

Figure 24 - V414 Sct - g mag vs Phase (308 ± 32 days period - weighted average)

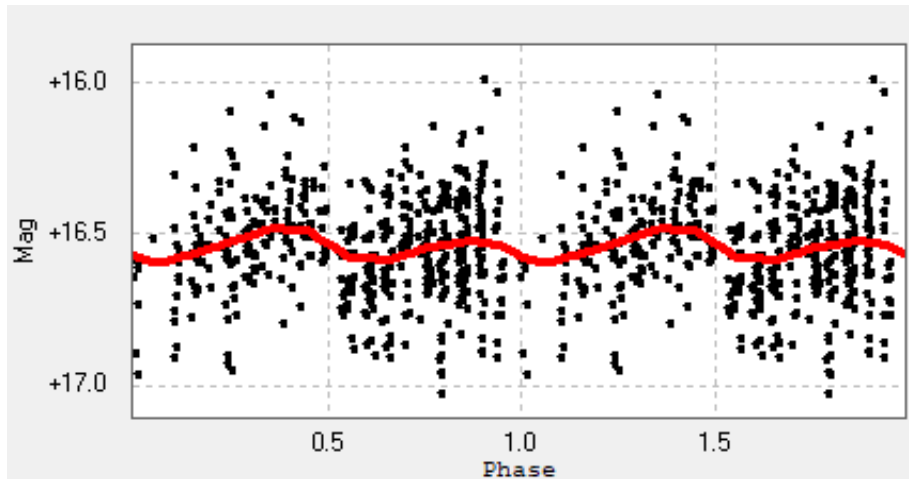

Figure 25 - V414 Sct - g mag vs Phase (592 ± 77 days period - ANOVA)

**V415 Sct**

This star is identified with the infrared counterpart 2MASS J18301342-1425191 and Gaia DR2/EDR3 source ID 4104132140880936576.

The SIMBAD database reports for this variable a Mira type in accordance with the original study, which determined a period of 303 days.

This classification is consistent with the Gaia DR2 color index Bp-Rp = 5.895 and absolute magnitude $M_G$ = 3.45 (-0.71; +0.54), which place this object in the LPV group of Figure 1.

It is classified as a large amplitude variable in Gaia DR2, with an amplitude $\Delta G$ = 0.32 mag, and the Bochum Galactic Disk Survey II associates this star to source ID GDS J1830133-142518, with a median light curve magnitude r = 16.91, i = 14.10 and an amplitude of 4.29 mag.

The ASAS-SN Catalog of Variable Stars II refers to this star as ASASSN-V J183013.34-142520.0 and classifies it as a Semiregular, with a mean magnitude V = 15.87, an amplitude of 1.24 mag and a period of 314 days.

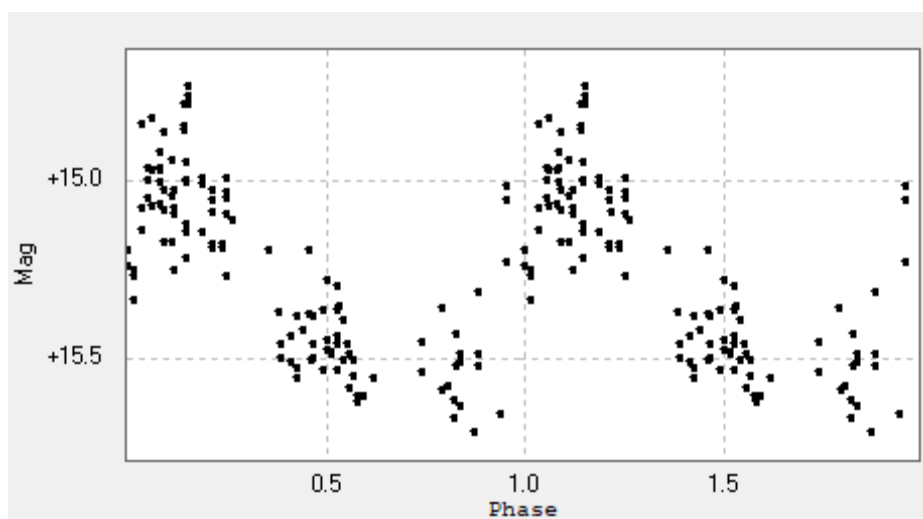

Figure 26 - V415 Sct - V mag vs Phase (75.8 ± 0.6 days period - weighted average)

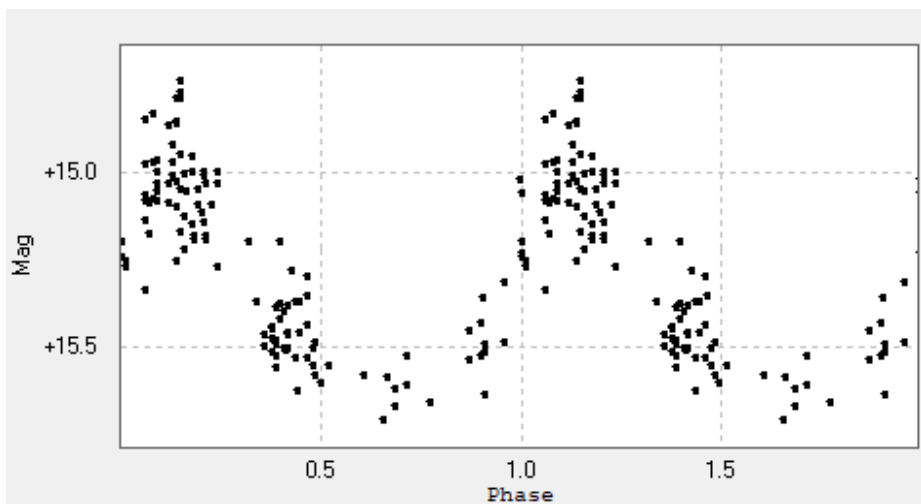

Figure 27 - V415 Sct - V mag vs Phase (100.3 ± 1.1 days period - weighted average)

We performed a period analysis, based on 121 and 172 valid observations available from ASAS-SN in the V filter and g, covering a time span of 1283 and 1079 days, respectively.
Applying the Lomb-Scargle and ANOVA methods to the set of observations in the filter V, we highlighted two potential solutions at the weighted average period of 75.8 ± 0.6 (Figure 26) and 100.3 ± 1.1 (Figure 27) days. No reliable solutions were found with the filter g.
We identified a maximum at epoch 2457988 ± 4 HJD.

**V416 Sct**
This star is identified with the infrared counterpart 2MASS J18301476-1421339 and Gaia DR2/EDR3 source ID 4104133519686449920. The SIMBAD database reports this variable as a Mira, in accordance with the original study, which determined a period of 300 days. This classification is consistent with the Gaia DR2 color index Bp-Rp = 5.925 and absolute magnitude $M_G$ = 0.43 (-0.98; +1.01), which place this object in the LPV group of Figure 1.
The Gaia DR2 classifies this variable as a LPV candidate with a period of 314 ± 44 days and an amplitude of 0.58 mag.
The ASAS-SN Catalog of Variable Stars II refers to this star as ASASSN-V J183014.76-142133.4 and classifies it as a Semiregular, with a mean magnitude V = 16.48, an amplitude of 0.88 mag and a period of 305 days.
We did not perform any analysis of the ASAS-SN photometric data due to insufficient valid measurements available.

**V417 Sct**
This star is identified with the infrared counterpart 2MASS J18301572-1431265 and Gaia DR2/EDR3 source ID 4104126815120395776. The SIMBAD database reports this variable as a Mira, in accordance with the original study, which determined a period of 407 days. This classification is consistent with the Gaia DR2 color index Bp-Rp = 6.229 and absolute magnitude $M_G$ = 0.56 (-1.30; +1.24), which place this object in the LPV group of Figure 1.

It is classified as a large amplitude variable in Gaia DR2, with an amplitude ΔG = 0.26 mag, and the Bochum Galactic Disk Survey II associates this star to source ID GDS J1830156-143126, with a median light curve magnitude r = 15.37, i = 13.17 and an amplitude of 2.66 mag.

The ASAS-SN Catalog of Variable Stars II associates V417 Sct to source ASASSN-V J183015.71-143126.4 and classifies it as a Semiregular, with a mean magnitude V = 16.09, an amplitude of 1.92 mag and a period of 435 days.

We performed a period analysis, based on 91 observations available from ASAS-SN in the V filter, covering a time span of 853 days. Applying the Lomb-Scargle and ANOVA methods we highlighted a potential period with a weighted average value of 118 ± 4 days (Figure 28). No reliable solution was found around 300 days. Not enough valid observations were available in the g filter. We identified a maximum at epoch 2457610 ± 3 HJD.

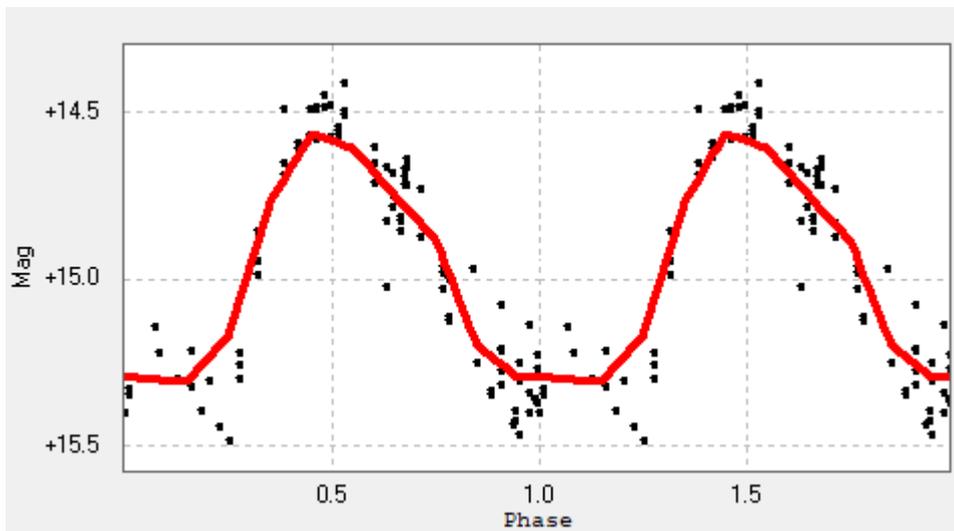

Figure 28 - V417 Sct - V mag vs Phase (118 ± 4 days period - weighted average)

**V418 Sct**

This star is identified with the infrared counterpart 2MASS J18302897-1421353 and Gaia DR2/EDR3 source ID 4104156197098012288. The SIMBAD database reports this variable as a Mira, in accordance with the original study, which determined a period of 411 days. This classification is consistent with the Gaia DR2 color index Bp-Rp = 6.159 and absolute magnitude $M_G$ = -0.38 (-0.79; +0.81), which place this object in the LPV group of Figure 1.

It is classified as a large amplitude variable in Gaia DR2, with an amplitude ΔG = 0.50 mag, and the Bochum Galactic Disk Survey II associates this star to source ID GDS J1830289-142135, with a median light curve magnitude i = 13.87 and an amplitude of 2.27 mag.

The ASAS-SN Catalog of Variable Stars II refers to this star as ASASSN-V J183028.96-142135.5 and classifies it as a Semiregular, with a mean magnitude V = 16.95, an amplitude of 0.54 mag and a period of 18 days.

We did not perform any analysis of the ASAS-SN photometric data due to insufficient valid measurements available.

**V419 Sct**

This star is identified with the infrared counterpart 2MASS J18303629-1416248 and Gaia DR2/EDR3 source ID 4104157915085065088. The SIMBAD database reports this variable as a Mira based on its preliminary classification (Maffei, 1975), that is superseded by the original study, which classified this variable as a Semiregular (SRa) and determined a period of 359 days. This classification is consistent with the Gaia DR2 color index Bp-Rp = 5.811 and absolute magnitude $M_G$ = 2.35 (-1.20; +1.16), which place this object in the LPV group of Figure 1.

It is classified as a large amplitude variable in Gaia DR2, with an amplitude $\Delta G$ = 0.32 mag.

No object is identified as a variable star by the ASAS-SN Catalog of Variable Stars II within 10" of the Gaia EDR3 equatorial coordinates of source ID 4104157915085065088.

We performed a period analysis, based on 707 and 1131 valid observations available from ASAS-SN in the V filter an g, covering a time span of 1337 and 1125 days, respectively. Applying the ANOVA and Lomb-Scargle methods to both set of observations, we highlighted a potential period of 29.6 ± 0.1 days (Figure 29). We found also other solutions around 180 and greater than 300 days, but because they are not statistically valid in all combination of methods and filters we considered these longer values not reliable. We also identified a maximum at epoch 2458297 ± 4 HJD.

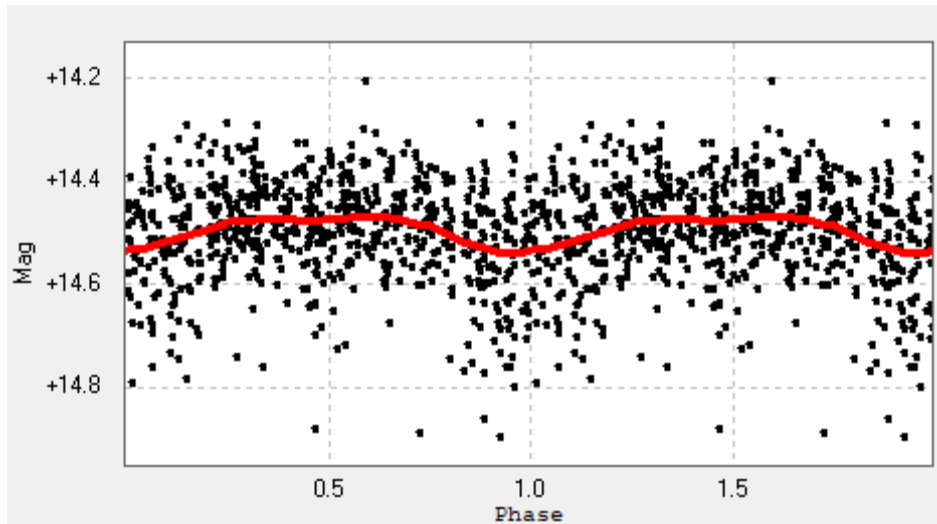

Figure 29 - V419 Sct - V mag vs Phase (29.6 ± 0.1 days period - weighted average)

**V420 Sct**

This star is identified with the infrared counterpart 2MASS J18304816-1452459 and Gaia DR2/EDR3 source ID 4104053633283796736. The SIMBAD database reports this variable as a Mira, in accordance with the original study, which determined a period of 261 days.

This classification is consistent with the Gaia DR2 color index Bp-Rp = 5.515 and absolute magnitude $M_G$ = 0.33 (-1.15; +0.99), which place this object in the LPV group of Figure 1.

The Gaia DR2 classifies this variable as a LPV candidate with a period of 274 ± 34 days and an amplitude of 0.47 mag.

The Bochum Galactic Disk Survey II refers to this variable as source ID GDS J1830481-145246, with a median light curve magnitude r = 15.32, i = 12.69 and an amplitude of 3.39 mag.

The ASAS-SN Catalog of Variable Stars II refers to this star as ASASSN-V J183048.13-145245.5 and classifies it as a Semiregular, with a mean magnitude V = 14.58, an amplitude of 1.62 mag and a period of 257 days.

We performed a period analysis, based on 693 and 1151 valid observations available from ASAS-SN in the V filter an g, covering a time span of 1337 and 1112 days, respectively. Applying the ANOVA and Lomb-Scargle methods to both set of observations, we determined a potential period, as weighted average of all solutions, of 259 ± 2 days (Figure 30). We also identified a maximum at epoch 2458370 ± 2 HJD.

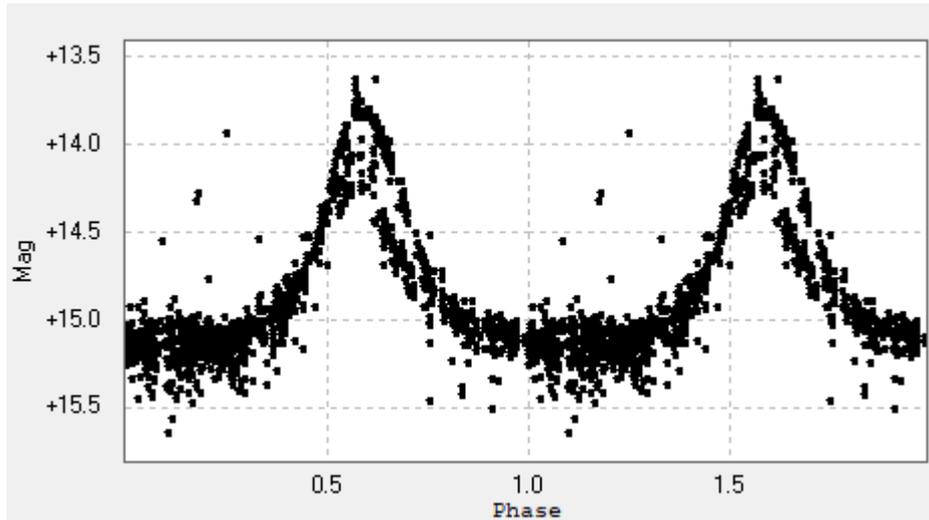

Figure 30 - V420 Sct - V mag vs Phase (259 ± 2 days period - weighted average)

**V421 Sct**

This star is identified with the infrared counterpart 2MASS J18305093-1537289 and Gaia DR2/EDR3 source ID 4103167633028466816.

The original study classified this variable as an uncertain Mira and determined a period of 219.9 days. This classification is consistent with the Gaia DR2 color index Bp-Rp = 5.763 and absolute magnitude $M_G$ = 0.37 ± 0.95, which place this object in the LPV group of Figure 1.

It is classified as a large amplitude variable in Gaia DR2, with an amplitude ΔG = 0.52 mag.

The ASAS-SN Catalog of Variable Stars II refers to this star as ASASSN-V J183050.93-153728.5 and classifies it as a Semiregular, with a mean magnitude V = 14.83, an amplitude of 0.79 mag and a period of 226 days.

We performed a period analysis, based on 693 and 1065 valid observations available from ASAS-SN in the V filter an g, covering a time span of 1335 and 1112 days, respectively. Applying the ANOVA and Lomb-Scargle methods to both set of observations, we determined a potential period, as weighted average of all solutions, of 221 ± 11 days (Figure 31). We also identified a maximum at epoch 2457156 ± 3 HJD.

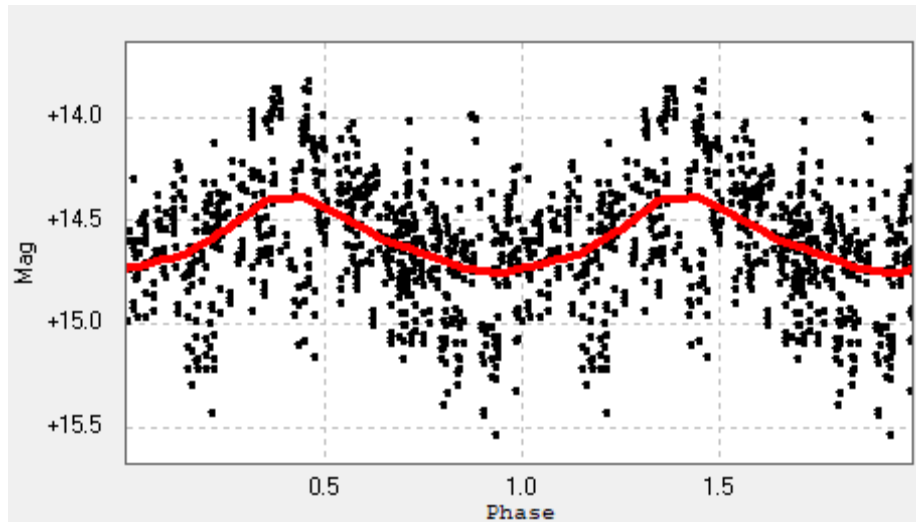

Figure 31 - V421 Sct - V mag vs Phase (221 ± 11 days period - weighted average)

**V422 Sct**
This star is identified with the infrared counterpart 2MASS J18312526-1518226 and Gaia DR2/EDR3 source ID 4103183266710080896. The SIMBAD database reports this variable as a Mira, in accordance with the original study, which determined a period of 423 days. This classification is consistent with the Gaia DR2 color index Bp-Rp = 5.569 and absolute magnitude $M_G$ = 0.56 (-0.92; +0.95), which place this object in the LPV group of Figure 1.
It is classified as a large amplitude variable in Gaia DR2, with an amplitude $\Delta G$ = 0.87 mag, and the Bochum Galactic Disk Survey II associates this star to source ID GDS J1831252-151822, with a median light curve magnitude i = 15.17 and an amplitude of 2.21 mag.
No object is identified as a variable star by the ASAS-SN Catalog of Variable Stars II within 10" of the Gaia EDR3 equatorial coordinates of source ID 4103183266710080896.
We did not perform any analysis of the ASAS-SN photometric data due to insufficient valid measurements available.

**V423 Sct**
This star is identified with the infrared counterpart 2MASS J18312523-1443501 and Gaia DR2/EDR3 source ID 4104062463737773568. The SIMBAD database reports this variable as a Mira Candidate, in accordance with the original study, which determined a period of 449 days. This classification is consistent with the Gaia DR2 color index Bp-Rp = 5.321 and absolute magnitude $M_G$ = 2.07 (-1.30; +1.18), which place this object in the LPV group of Figure 1.
It is classified as a large amplitude variable in Gaia DR2, with an amplitude $\Delta G$ =0.61, and reported as a WISE J183125.24-144350.2 YSO candidate.
The Bochum Galactic Disk Survey II refers to this variable as source ID GDS J1831252-144350, with a median light curve magnitude i = 14.38 and an amplitude of 2.92 mag.
No object is identified as a variable star by the ASAS-SN Catalog of Variable Stars II within 10" of the Gaia EDR3 equatorial coordinates of source ID 4104062463737773568.

We did not perform any analysis of the ASAS-SN photometric data due to insufficient valid measurements available.

**V424 Sct**

This star is identified with the infrared counterpart 2MASS J18313627-1526581 and Gaia DR2/EDR3 source ID 4103177047597293696. The SIMBAD database reports this variable as a Mira, in accordance with the original study, which determined a period of 474 days. This classification is consistent with the Gaia DR2 color index Bp-Rp = 5.271 and absolute magnitude $M_G$ = 1.30 (-1.10; +1.13), which place this object in the LPV group of Figure 1.

It is classified as a large amplitude variable in Gaia DR2, with an amplitude $\Delta G$ =0.61, and reported as a WISE YSO J183136.29-152658.0 candidate.

The Bochum Galactic Disk Survey II refers to this variable as source ID GDS J1831362-152658, with a median light curve magnitude i = 14.89 and an amplitude of 2.64 mag.

The ASAS-SN Catalog of Variable Stars II refers to this star as ASASSN-V J183136.27-152658.1 and classifies it as a Semiregular, with a mean magnitude V = 16.66, an amplitude of 0.48 mag and a period of 16 days.

We did not perform any analysis of the ASAS-SN photometric data due to insufficient valid measurements available.

**V425 Sct**

This star is identified with the infrared counterpart 2MASS J18314233-1512142 and Gaia DR2/EDR3 source ID 44103289674449169408. The SIMBAD database reports this variable as a Mira, in accordance with the original study, which determined a period of 445 days. This classification is consistent with the Gaia DR2 color index Bp-Rp = 7.014 and absolute magnitude $M_G$ = 1.47 (-1.26; +1.28), which place this object in the LPV group of Figure 1.

It is classified as a large amplitude variable in Gaia DR2, with an amplitude $\Delta G$ =0.51, and the Bochum Galactic Disk Survey II refers to this variable as source ID GDS J1831423-151214, with a median light curve magnitude i = 15.39 and an amplitude of 3.66 mag.

The ASAS-SN Catalog of Variable Stars II associates V425 Sct to source ID ASASSN-V J183142.34-151214.2, and classifies it as a Semiregular, with a mean magnitude V = 17.11, an amplitude of 0.78 mag and a period of 489 days.

We did not perform any analysis of the ASAS-SN photometric data due to insufficient valid measurements available.

## 3. Conclusions

The assessment of data available from public astronomical databases and referenced papers, and the analysis we performed on light curves and periods, refine the original classification, or suggest a revision of the type and/or the period for several of the 55 variables, part of the Scutum constellation of IBVS 985.

For all stars, the variability is confirmed.

For the following 17 stars, our analysis provides refinements or solutions for the type and/or period significantly different from existing studies: NSV 10736, NSV 10832, NSV 10899, V374 Sct, V378 Sct, V381 Sct, V383 Sct, V388 Sct, V392 Sct, V393 Sct, V400 Sct, V406 Sct, V408 Sct, V414 Sct, V415 Sct, V417 Sct, V419 Sct.

Our assessment also identifies 12 cases for which characteristics available from ASAS-SN photometric database are significantly different from the original classification. The variables V376 Sct, V379 Sct, V380 Sct, V383 Sct, V386 Sct, V392 Sct, V395 Sct, V404 Sct, V405 Sct, V418 Sct, V424 Sct, V425 Sct are all classified as Mira in the original study, with periods in the range from 288 to 590 days, whilst ASAS-SN database classifies them as Semiregular with shorter periods in the range from 5 to 42 days.

We noted this discrepancy in two previous works: La Rocca et al. 2020, where original light curves were not available, and Aglì et al. 2021. Due to the good quality of the available original Mira light curves we deem that this discrepancy is due to the rules applied by the ASAS-SN system for the classification of the types and the period determination.

Our assessment also highlights that for 3 stars, V382 Sct, V402 Sct and V406 Sct, incorrect cross-reference names are reported by ASAS-SN Catalog of Variable Stars.

Table 7 provides a summary of the main results of this work.

Table 7       Summary of the main results of this work

| (Maffei and Tosti, 2013) | | | This assessment | | |
|---|---|---|---|---|---|
| **Variable** | **Type** | **Period (d)** | **Type** | **Period (d)** | **Remarks** |
| NSV 10736 | M | 507.0 | Probable YSO (our analysis)<br><br>SR (ASAS-SN) | Not applicable (our analysis)<br><br>627 (ASAS-SN) | NSV 10736 = V478 Sct |
| NSV 10832 | E | --- | LPV (our analysis)<br><br>SR (GDS II) | --- (our analysis)<br><br>--- (GDS II) | --- |
| NSV 10848 | Nova? | --- | --- (our analysis)<br><br>Probable Nova (Nesci 2018) | --- | Not consistent finding charts make identification uncertain (our analysis) |
| NSV 10899 | I | --- | LPV (our analysis)<br><br>Candidate YSO (WISE) | 185 ± 8 (our analysis)<br><br>Not applicable (WISE) | --- |

| (Maffei and Tosti, 2013) | | | This assessment | | |
|---|---|---|---|---|---|
| **Variable** | **Type** | **Period (d)** | **Type** | **Period (d)** | **Remarks** |
| V374 Sct | E | 408.0 | LPV (our analysis) | 29.6 ± 0.3 (our analysis) | --- |
| | | | SR (ASAS-SN) | 14 (ASAS-SN) | |
| V375 Sct | M | 495.0 | --- (our analysis) | --- | --- |
| V376 Sct | M | 590.0 | LPV (our analysis) | 591 ± 81 (our analysis) | --- |
| | | | SR (ASAS-SN) | 5 (ASAS-SN) | |
| V377 Sct | M | 522.0 | LPV (our analysis) | --- (our analysis) | --- |
| | | | LPV candidate (Gaia DR2) | 510 ± 134 (Gaia DR2) | |
| V378 Sct | M | 435.0 | LPV (our analysis) | 180 ± 9 (our analysis) | --- |
| | | | LPV candidate (Gaia DR2) | 468 ± 61 (Gaia DR2) | |
| | | | SR (ASAS-SN) | 164 (ASAS-SN) | |
| V379 Sct | M | 413.0 | LPV (our analysis) | --- (our analysis) | --- |
| | | | SR (ASAS-SN) | 22 (ASAS-SN) | |
| V380 Sct | M | 447.0 | LPV (our analysis) | 29.5 ± 0.2 (our analysis) | Brighter Gaia EDR3 source ID 4104201998549191936 (G = 14.820) is located 4.8 arcsec southwest of V380 Sct |
| | | | LPV candidate (Gaia DR2) | 388 ± 54 (Gaia DR2) | |
| | | | SR (ASAS-SN) | 13 (ASAS-SN) | |
| V381 Sct | SRa | 417.0 | LPV (our analysis) | 28.5 ± 0.2 183 ± 9 (our analysis) | --- |
| | | | SR (ASAS-SN) | 171 (ASAS-SN) | |
| V382 Sct | E: | 592: | LPV (our analysis) | --- (our analysis) | ASAS-SN cross reference to Gaia EDR3 ID 4104199043579502592 is incorrect. ASAS-SN light curve mean magnitude V = 9.51 is not consistent with V382 Sct brightness (our analysis) |
| | | | LPV candidate (Gaia DR2) | 577 ± 101 (Gaia DR2) | |

| (Maffei and Tosti, 2013) | | | This assessment | | |
|---|---|---|---|---|---|
| **Variable** | **Type** | **Period (d)** | **Type** | **Period (d)** | **Remarks** |
| V383 Sct | M | 289.0 | LPV (our analysis) | 13.07 ± 0.03 (our analysis) | --- |
| | | | SR (ASAS-SN) | 13 (ASAS-SN) | |
| V384 Sct | M: | 454.0 | LPV (our analysis) | --- (our analysis) | --- |
| | | | Candidate YSO (WISE) | Not applicable (WISE) | |
| V385 Sct | M | 397.0 | LPV (our analysis) | 394 ± 3 (our analysis) | Maximum = 2458014 ± 2 HJD (our analysis) |
| | | | LPV Candidate (Gaia DR2) | 388 ± 22 (Gaia DR2) | |
| | | | M (ASAS-SN) | 404 (ASAS-SN) | |
| V386 Sct | M | 426.0 | LPV (our analysis) | No reliable period found (our analysis) | --- |
| | | | SR (ASAS-SN) | 16 (ASAS-SN) | |
| V387 Sct | M | 379.0 | LPV (our analysis) | --- (our analysis) | --- |
| V388 Sct | M | 379.0 | LPV (our analysis) | 196 ± 7 (our analysis) | --- |
| V390 Sct | M | 319.0 | LPV (our analysis) | --- (our analysis) | --- |
| | | | Candidate YSO (WISE) | Not applicable (WISE) | |
| V391 Sct | IS: | --- | Not LPV (our analysis) | ---- (our analysis) | --- |
| | | | U Gem: (Cataclysmic. Var. Catalog) | Not applicable (Cataclysmic. Var. Catal.) | |
| | | | R CrB (Tisserand) | Not applicable (Tisserand) | |
| | | | R CrB (ASAS-SN) | 486 (ASAS-SN) | |
| V392 Sct | M | 480.0 | Probable YSO (our analysis) | Not applicable (our analysis) | --- |
| | | | Candidate YSO (WISE) | Not applicable (WISE) | |
| | | | SR (ASAS-SN) | 42 (ASAS-SN) | |

| (Maffei and Tosti, 2013) | | | This assessment | | |
|---|---|---|---|---|---|
| **Variable** | **Type** | **Period (d)** | **Type** | **Period (d)** | **Remarks** |
| V393 Sct | M | 381.5 | LPV (our analysis) | 338 ± 20 (our analysis) | Maximum at epoch 2458596 ± 2 HJD (our analysis) |
| | | | SR (ASAS-SN) | 388 (ASAS-SN) | |
| V394 Sct | M | 314.0 | LPV (our analysis) | 323 ± 18 (our analysis) | Maximum at epoch 2458299 ± 4 HJD (our analysis) |
| | | | SR (ASAS-SN) | 332 (ASAS-SN) | |
| V395 Sct | M | 310.0 | LPV (our analysis) | 314 ± 22 (our analysis) | --- |
| | | | LPV Candidate (Gaia DR2) | 306 ± 10 (Gaia DR2) | |
| | | | SR (ASAS-SN) | 24 (ASAS-SN) | |
| V396 Sct | M | 398.0 | LPV (our analysis) | Not reliable period found (our analysis) | --- |
| | | | Candidate YSO (WISE) | Not applicable (WISE) | |
| | | | SR (ASAS-SN) | 5 (ASAS-SN) | |
| V397 Sct | M | 312.0 | LPV (our analysis) | 306 ± 16 (our analysis) | --- |
| | | | Candidate YSO (WISE) | Not applicable (WISE) | |
| | | | SR (ASAS-SN) | 168 (ASAS-SN) | |
| V398 Sct | M | 368.0 | LPV (our analysis) | --- (our analysis) | --- |
| V399 Sct | M | 317.0 | LPV (our analysis) | --- (our analysis) | --- |
| V400 Sct | M | 295.0 | LPV (our analysis) | 113 ± 2 318 ± 13 (our analysis) | --- |
| | | | SR (ASAS-SN) | 297 (ASAS-SN) | |
| V401 Sct | M | 474.0 | LPV (our analysis) | --- (our analysis) | --- |
| | | | LPV Candidate (Gaia DR2) | 432 ± 92 (Gaia DR2) | |

| (Maffei and Tosti, 2013) | | | This assessment | | |
|---|---|---|---|---|---|
| **Variable** | **Type** | **Period (d)** | **Type** | **Period (d)** | **Remarks** |
| V402 Sct | M | 312.0 | LPV (our analysis) | 314 ± 6 (our analysis) | ASAS-SN cross reference to Gaia EDR3 ID 4104026660847518 9764 is probably incorrect (our analysis) |
| | | | LPV Candidate (Gaia DR2) | 432 ± 92 (Gaia DR2) | |
| | | | SR (ASAS-SN) | 297 (ASAS-SN) | Maximum at epoch 2459041 ± 2 HJD (our analysis) |
| V403 Sct | M | 337.0 | LPV (our analysis) | --- (our analysis) | --- |
| V404 Sct | M | 288.0 | LPV (our analysis) | --- (our analysis) | --- |
| | | | Candidate YSO (WISE) | Not applicable (WISE) | |
| | | | SR (ASAS-SN) | 18 (ASAS-SN) | |
| V405 Sct | M | 307.0 | LPV (our analysis) | No reliable period found (our analysis) | --- |
| | | | Candidate YSO (WISE) | Not applicable (WISE) | |
| | | | SR (ASAS-SN) | 42 (ASAS-SN) | |
| V406 Sct | M: | 447.0 | LPV (our analysis) | 192 ± 16 291 ± 29 455 ± 66 (our analysis) | ASAS-SN cross reference to Gaia EDR3 source ID 4103139011331347200 is incorrect (our analysis) |
| | | | LPV Candidate (Gaia DR2) | 459 ± 31 (Gaia DR2) | Maximum at epoch 2457941 ± 10 HJD (our analysis) |
| | | | YSO (ASAS-SN) | N/A (ASAS-SN) | |
| V407 Sct | M | 238.5 | LPV (our analysis) | 234 ± 1 (our analysis) | Maximum at epoch 2459026 ± 4 HJD (our analysis) |
| | | | LPV Candidate (Gaia DR2) | 231 ± 5 (Gaia DR2) | |
| | | | M (ASAS-SN) | 178 (ASAS-SN) | |
| V408 Sct | M | 321.0 | Probable YSO (our analysis) | 29.4 ± 0.2 (our analysis) | Maximum at epoch 2458302 ± 2 HJD (our analysis) |
| | | | Candidate YSO (WISE) | Not applicable (WISE) | |
| | | | SR (ASAS-SN) | 166 (ASAS-SN) | |

| (Maffei and Tosti, 2013) | | | This assessment | | |
|---|---|---|---|---|---|
| **Variable** | **Type** | **Period (d)** | **Type** | **Period (d)** | **Remarks** |
| V409 Sct | M | 469.0 | LPV (our analysis)<br><br>SR (ASAS-SN) | No reliable period found (our analysis)<br><br>342 (ASAS-SN) | Incorrect SIMBAD cross reference to 2MASS J18293968-1359367 and Gaia DR2/EDR3 4104366341197354624 (our analysis) |
| V410 Sct | M | 327.0 | LPV (our analysis) | --- (our analysis) | --- |
| V411 Sct | SR | 457.0 | LPV (our analysis) | --- (our analysis) | --- |
| V412 Sct | M | 408.0 | LPV (our analysis) | --- (our analysis) | --- |
| V413 Sct | M | 333.0 | LPV (our analysis)<br><br>L (ASAS-SN) | No reliable period found (our analysis)<br><br>--- (ASAS-SN) | --- |
| V414 Sct | M | 529.0 | --- (our analysis)<br><br>Candidate YSO (WISE)<br><br>VAR (ASAS-SN) | 308 ± 32<br>592 ± 77 (our analysis)<br><br>N/A (WISE)<br><br>No period defined (ASAS-SN) | Brighter and bluer Gaia DR2/EDR3 ID 4103185633161346560 within 2 arcsec<br><br>Maximum at epoch 2458952 ± 3 HJD (our analysis) |
| V415 Sct | M | 303.0 | LPV (our analysis)<br><br>SR (ASAS-SN) | 75.8 ± 0.6<br>100.3 ± 1.1 (our analysis)<br><br>314 (ASAS-SN) | Maximum at epoch 2457988 ± 4 HJD (our analysis) |
| V416 Sct | M | 300.0 | LPV (our analysis)<br><br>LPV Candidate (Gaia DR2)<br><br>SR (ASAS-SN) | --- (our analysis)<br><br>314 ± 44 (Gaia DR2)<br><br>305 (ASAS-SN) | --- |
| V417 Sct | M | 407.0 | LPV (our analysis)<br><br>SR (ASAS-SN) | 118 ± 4 (our analysis)<br><br>435 (ASAS-SN) | Maximum at epoch 2457610 ± 3 HJD (our analysis) |
| V418 Sct | M | 411.0 | LPV (our analysis)<br><br>SR (ASAS-SN) | --- (our analysis)<br><br>18 (ASAS-SN) | --- |
| V419 Sct | SRa | 359.0 | LPV (our analysis) | 29.6 ± 0.1 (our analysis) | Maximum at epoch 2458297 ± 4 HJD (our analysis) |

| | (Maffei and Tosti, 2013) | | This assessment | | |
|---|---|---|---|---|---|
| **Variable** | **Type** | **Period (d)** | **Type** | **Period (d)** | **Remarks** |
| V420 Sct | M | 261.0 | LPV (our analysis) | 259 ± 2 (our analysis) | Maximum at epoch 2458370 ± 2 HJD (our analysis) |
| | | | LPV Candidate (Gaia DR2) | 274 ± 34 (Gaia DR2) | |
| | | | SR (ASAS-SN) | 257 (ASAS-SN) | |
| V421 Sct | M: | 219.9 | LPV (our analysis) | 221 ± 11 (our analysis) | Maximum at epoch 2457156 ± 3 HJD (our analysis) |
| | | | SR (ASAS-SN) | 226 (ASAS-SN) | |
| V422 Sct | M | 423.0 | LPV (our analysis) | --- (our analysis) | --- |
| V423 Sct | M | 449.0 | LPV (our analysis) | --- (our analysis) | --- |
| | | | Candidate YSO (WISE) | Not applicable (WISE) | |
| V424 Sct | M | 474.0 | LPV (our analysis) | --- (our analysis) | --- |
| | | | Candidate YSO (WISE) | Not applicable (WISE) | |
| | | | SR (ASAS-SN) | 16 (ASAS-SN) | |
| V425 Sct | M | 445.0 | LPV (our analysis) | --- (our analysis) | --- |
| | | | SR (ASAS-SN) | 489 (ASAS-SN) | |

**Acknowledgements**

- This activity has made use of the SIMBAD database, operated at CDS, Strasbourg, France.
- This work has made use of data from the European Space Agency (ESA) mission Gaia (https://www.cosmos.esa.int/gaia), processed by the Gaia Data Processing and Analysis Consortium (DPAC, https://www.cosmos.esa.int/web/gaia/dpac/consortium). Funding for the DPAC has been provided by national institutions, in particular the institutions participating in the Gaia Multilateral Agreement.
- We acknowledge with thanks the variable star observations from the *AAVSO International Database* contributed by observers worldwide and used in this research.
- This work was carried out in the context of educational and training activities provided by Italian law 'Percorsi per le Competenze Trasversali e l'Orientamento', December 30th, 2018, n.145, Art.1.